\documentclass[graybox]{./styles/svmult}

\usepackage{mathptmx}       
\usepackage{helvet}         
\usepackage{courier}        
\usepackage{makeidx}         
\usepackage{graphicx}        
\usepackage{multicol}        
\usepackage[bottom]{footmisc}

\makeindex             

\usepackage[nolist]{acronym}
\usepackage{todonotes}

\usepackage{calc}
\usepackage{graphicx}
\usepackage{amsmath}
\usepackage{amssymb}
\usepackage{relsize}
\usepackage{multirow}
\usepackage{rotating}
\usepackage{bm}
\usepackage{url}
\usepackage{graphicx}
\usepackage{multicol}
\usetikzlibrary{calc}
\usepackage{pgfplots}

\usetikzlibrary{arrows, decorations.markings}
\tikzset{thicker line small arrows m/.style args={#1in#2}{
   draw=#2,
   solid,
   line width=#1,
   shorten >=1mm,
   decoration={
     markings,
     mark=at position 1.0 with {\arrow[fill=#2,thin]{triangle 90}}
    },
   postaction={decorate}
  }}

\definecolor{lightblue}{rgb}{0.145,0.6666,1}
\definecolor{imperialnavy}{RGB}{0,33,71}
\definecolor{imperialblue}{RGB}{0,62,116}
\definecolor{imperialgrey}{RGB}{235,238,238}
\definecolor{imperialcoolgrey}{RGB}{157,157,157}
\definecolor{grey52}{RGB}{52,52,52}
\definecolor{color1}{RGB}{0,62,116}
\definecolor{color2}{RGB}{152,152,152}
\definecolor{color3}{RGB}{52,52,52}
\definecolor{color4}{RGB}{100,100,100}

\newcommand{\prt}[1]{\left(#1\right)}

\newcommand{\abs}[1]{\left|#1\right|}
\newcommand{\norm}[1]{\left\|#1\right\|}
\newcommand{\set}[1]{\left\{#1\right\}}

\DeclareMathOperator*{\argmin}{arg\,min}

\newcommand{\der}[2]{\frac{\mathrm{d} #1}{\mathrm{d} #2}}

\usepackage{multicol}
\usepackage{mathtools}
\usepackage{nicefrac}

\usepackage{varioref}
\usepackage{hyperref}
\usepackage[capitalise]{cleveref}
\usepackage{xcolor}
\hypersetup{
	colorlinks=true,
	linkcolor=blue,
	citecolor=blue,
	urlcolor=blue,
	linktocpage
}

\newcommand{\defeq}{\vcentcolon=}

\usepackage{lipsum}
\usepackage{rotating}

\usepackage{breakurl}
\begin{document}
\title*{Pedestrian Models based on Rational Behaviour}
\author{Rafael Bailo, Jos\'{e} A. Carrillo, Pierre Degond}
\institute{
R. Bailo $\cdot$ J. A. Carrillo $\cdot$ P. Degond
\at Department of Mathematics, Imperial College London, London, United Kingdom\\
email:
\href{mailto:r.bailo@imperial.ac.uk}{r.bailo@imperial.ac.uk};
\href{mailto:carrillo@imperial.ac.uk}{carrillo@imperial.ac.uk};
\href{mailto:p.degond@imperial.ac.uk}{p.degond@imperial.ac.uk}
}
%
%

\maketitle
\abstract{
 Following the paradigm set by attraction-repulsion-alignment schemes, a myriad of individual based models have been proposed to calculate the evolution of abstract agents. While the emergent features of many agent systems have been described astonishingly well with force-based models, this is not the case for pedestrians. Many of the classical schemes have failed to capture the fine detail of crowd dynamics, and it is unlikely that a purely mechanical model will succeed. As a response to the mechanistic literature, we will consider a model for pedestrian dynamics that attempts to reproduce the rational behaviour of individual agents through the means of anticipation. Each pedestrian undergoes a two-step time evolution based on a perception stage and a decision stage. We will discuss the validity of this game theoretical based model in regimes with varying degrees of congestion, ultimately presenting a correction to the mechanistic model in order to achieve realistic high-density dynamics.
}

\section{Introduction}

The behaviour of humans moving in crowds was studied early on from the engineering perspective \cite{Fruin1971, Hankin1958, Older1968}. These works were based on the observation of crowds, either directly or through photographs and film, and ultimately aimed to provide planning guidelines and construction directives such as \cite{HCM1985,HCM2000}. These studies have always had an economic concern, but most importantly a safety outlook, as a good understanding of crowd dynamics can help prevent the injuries and deaths which derive from the confluence of inordinate numbers of people in places that are not prepared for such occupancy \cite{Helbing2005, Helbing2000}, for instance, as a result of a popular sporting event \cite{Gill2004} or a concert \cite{Johnson1987}. A review of incidents of this type can be found in \cite{Ngai2009}.

The walking behaviour of humans is extremely complex and not easy to capture in models. The observed phenomena in crowds is often unintuitive, for instance, small obstacles in the way of an exit can serve to stabilise the flow and make traffic more efficient \cite{Helbing2005}; when it comes to emergencies, it is found that evacuation is safer and more efficient at lower speeds \cite{Helbing2000}.

Extensive experimental work in the area has focused in numerous aspects of the dynamics of pedestrians, such as the behaviour of agents around bottlenecks \cite{Daamen2003, Daamen2003a, Kretz2006a}, intersections \cite{Helbing2005}, in counterflows \cite{Kretz2006}, following behaviours \cite{Lemercier2012}, cluster formation \cite{Moussaid2012}, the effects of fatigue \cite{Luo2016} and the empirical relation between crowd density and walking speed \cite{Seyfried2005} (known as the fundamental diagram). These studies have led to some understanding of the emergent features, those that arise not from the actions of any particular individual but rather as a result of the interactions of the collective. Stop-and-go waves \cite{Helbing2007}, lane formation \cite{Helbing2001, Moussaid2009} (somewhat a human counterpart to flocking \cite{Reynolds1987}), the crowding behaviour around bottlenecks \cite{Daamen2003, Helbing2000} and the fluid-like properties (shockwaves, turbulence) displayed by extremely dense crowds \cite{Helbing2007a, Helbing2007} are just some examples of the rich and subtle properties of the dynamics. Some of these features have been successfully reproduced in models, for instance, lane formation in \cite{Helbing1995, Moussaid2011}. Some, such as the fundamental diagram \cite{Weidmann1993}, have evaded many modelling attempts and are still the subject of avid debate despite substantial experimental work in the area \cite{Appert-Rolland2014, Johansson2010}, often requiring studies specific to particular configurations such as single direction flow \cite{Jelic2012, Seyfried2005}, behaviour around bottlenecks \cite{Kretz2006}, and assessment of the \textit{level of service} \cite{Mori1987, Polus1983}.

Pedestrians were early on modelled from a macroscopic perspective \cite{Henderson1971,Henderson1974,Lighthill1955,Lighthill1955a}, where only the features of the crowd as a whole (such as the pedestrian density, the flow through a corridor or the emergence of consensus \cite{Carrillo2010}) are of interest. Some of these models are prescribed directly \cite{Hughes2002, Hughes2003, Jiang2010}, and some derived from the kinetic point of view \cite{Bellomo2010a, Degond2013a, Helbing1992}. An overview can be found in \cite{Bellomo2011, Bellomo2011a}. Such models have been successful in providing an understanding of the large-scale behaviour, but provide no insight into the behaviour of individuals. The relation between the microscopic dynamics and the macroscopic scale in general is an active area of research. In humans, local effects have only been added to macroscopic models in some cases \cite{Carrillo2016}.

Many individual-based models have also been developed. A recent review can be found in \cite{Bellomo2011b}. These models are often based on alignment and force principles which follow in the reductionist philosophies of the social field \cite{Lewin1951} and the social forces \cite{Gibson1958}, which gave rise to models such as \cite{Helbing1995}. These microscopic models sometimes concern agents in the more abstract sense \cite{Braitenberg1984}. Often they deal with simple `animals', whether in general \cite{DOrsogna2006, Reynolds1987, Reynolds1999} or with a specific animal in mind \cite{Strombom2011, Strombom2014}. Sometimes they involve alignment processes \cite{Cucker2007, Cucker2007a}, and phase transitions were detected early on in works such as \cite{Vicsek1995}. These abstractions serve to study the natural emergence of self-organisation \cite{Cristiani2010, Ha2010} and swarming \cite{Carrillo2010a} as well as the methods to induce such feature when they do not occur spontaneously \cite{Borzi2015, Caponigro2013}. Furthermore, many agent models have been developed specifically for pedestrians: early force models \cite{Helbing1991, Helbing1995} and subsequent improvements \cite{Bajec2003, Moussaid2010}, models exploring self-organisation \cite{Helbing1998, Helbing1999, Lemercier2012}, as well as evacuation models \cite{Helbing2000}

While the force-based models are ubiquitous and relatively successful, there is a limit to what behaviours they can capture. Humans and their motion are not completely described by simple mechanistic models, as they fail to incorporate our rational behaviour. Not only are we capable of estimating the position and velocities of moving obstacles \cite{Cutting1995}, but we are able to assess the danger they pose to us \cite{Warren2004}. The literature in biology, psychology and neuroscience points to the existence of specialised neural mechanisms in the retina and the brain that enable pedestrians to detect potential obstacles and to assess the time until the collision with said obstacles occurs \cite{Hopkins2004, Schrater2000}. Those heuristics are then used by the agents to make quick, close-to-optimal adjustments to their trajectories in order to avoid possible collisions \cite{Batty1997, Gigerenzer2008}.

This work will discuss a model that attempts to replicate said rationality in order to realistically simulate the dynamics of pedestrians. Based on the principles of \cite{Moussaid2011} and the formulation of \cite{Degond2013}, the model fundamentally consists of a two-step evolution process: the first step involves the evaluation of heuristics of the environment and their use to estimate the proximity and dangerousness of encounters; the second concerns the decision-making process of each agent, which will involve an optimisation game in order to remain in motion towards a target while avoiding potential collisions. \cref{sec:rationalbehaviour} introduces the model in its original formulation, as well as a number of implementation alternatives. \cref{sec:highdensity} develops improvements and variations aimed to generalise the model to multiple situations and density regimes. \cref{sec:outlook} concludes the presentation of the model in a final assessment and presents the outlook of this work.

\section{A Model with Rational Behaviour}\label{sec:rationalbehaviour}
The pedestrian model of \cite{Degond2013,Moussaid2011} simulates the rational decision-making involved in the steering behaviour of agents. This model is an attempt to capture the complexity behind the steering behaviour of pedestrians. Collision avoidance on humans is an intricate conscious process, and any attempts to reproduce it through a purely mechanical set of rules can only achieve limited success.

The following section presents the formulation of the model as well as some implementation details. The model is conceptually fractioned into two steps: a \textit{perception stage} and a \textit{decision stage}. The perception stage comprises the use of visual stimuli to inform pedestrians of their environment, the surrounding obstacles (moving or not), and any potential collision. The decision stage encompasses the mechanisms through which each agent judges available paths and resolves to move in a specific direction.

\subsection{Perception Stage}
The perception stage is the first step towards collision avoidance in the model.
Pedestrians in this stage derive \textit{heuristics} about their environment based on the position and velocity of visible obstacles (objects or other pedestrians). The heuristics are first examined for encounters between only two agents. Afterwards, \textit{global} heuristics are considered.

\subsubsection{Pairwise Encounters}\label{sec:pairwiseencounters}

We shall first study a binary encounter, consisting of two pedestrians $i$ and $j$ which happen to approach each other as they move towards their targets. We assume that agent $i$ is aware of his own \textit{position} $x_i$ and \textit{velocity} $v_i$ and can also perceive $j$'s $x_j$ and $v_j$ accurately. This knowledge shall be used to derive \textit{heuristics} that will inform the decision making in the next section.

Throughout the text we will use the terms \textit{collision}, \textit{interaction} and \textit{encounter} interchangeably. All of these refer to a situation where two pedestrians approach enough to enter each other's \textit{personal space}; as a result they might be at risk of colliding, and the interaction must be resolved.

If the velocities of the agents are momentarily constant, their distance as a function of time can be expressed as:
\begin{align}
 d_{i,j}^2\prt{t}= & \norm{x_j+v_jt-x_i-v_it}^2,
 \\=&\label{eq:distance}\norm{v_j-v_i}^2\prt{t+\frac{\prt{x_j-x_i}\cdot\prt{v_j-v_i}}{\norm{v_j-v_i}^2}}^2
 \\\nonumber&+\norm{x_j-x_i}^2-\frac{\prt{\prt{x_j-x_i}\cdot\prt{v_j-v_i}}^2}{\norm{v_j-v_i}^2}.
\end{align}
The \textit{time to interaction of $i$ and $j$}, $\tau_{i,j}$, is the time that minimises $d_{i,j}$. This can be found by inspection of \eqref{eq:distance}, namely:
\begin{equation}\label{eq:tij}
 \tau_{i,j}=\argmin_{t\in\mathbb{R}}\set{d_{i,j}}=-\frac{\prt{x_j-x_i}\cdot\prt{v_j-v_i}}{\norm{v_j-v_i}^2}.
\end{equation}
Note that this time may be negative if the pedestrians are moving away from each other, i.e. $\prt{x_j-x_i}\cdot\prt{v_j-v_i}>0$; see \cref{fig:approachingpedestrians}.

\begin{figure}
 \sidecaption
 \begin{tikzpicture}[scale=0.57]
 \draw [dashed, color1] (0,0) -- (10,0);
 \path[thicker line small arrows m=1mm in color1] (0,0) -- (2.0,0) node[black,midway,above=0.1cm]{$v_i$};
 \fill[color1] (0,0) circle (0.50cm);
 \fill[white] (0,0) circle (0.25cm) node[black, below=0.5cm]{$x_i$};

 \draw [dashed, color2] (12,4) -- (6,-2);
 \path[thicker line small arrows m=1mm in color2] (12,4) -- ({12+2.0*cos(-135)},{4+2.0*sin(-135)}) node[black,midway,above left=0.1cm]{$v_j$};
 \fill[color2] (12,4) circle (0.50cm);
 \fill[white] (12,4) circle (0.25cm) node[black, below=0.5cm]{$x_j$};

 \fill[color1] (6,0) circle (0.1cm) node[black, above=0.055cm]{$p_{i,j}$};

 \fill[color2] (7,-1) circle (0.1cm) node[black, right=0.2cm]{$p_{j,i}$};

 \draw[|<->|, very thick, dotted, color3] (0,1.0) -- (6.02,1.0) node[black,draw=none,fill=none,midway,above=0.1cm] {
  \begin{tabular}{c}
   $0\leq D_{i,j}\leq L$ \\Distance to Interaction
  \end{tabular}
 };

 \draw[|<->|, very thick, dotted, color3] (5.69,-0.29) -- (6.71,-1.31) node[black,draw=none,fill=none,below=0.1cm,fill=white,opacity=.9,text opacity=1]
 {
  \begin{tabular}{c}
   $0\leq C_{i,j}\leq R$ \\Distance of Closest Approach
  \end{tabular}
 };
\end{tikzpicture}
 \caption{
  Dissection of a binary pedestrian encounter. Agent $i$ extrapolates his and $j$'s trajectories assuming their respective velocities will remain constant. This allows the estimation of \textit{distance of closest approach} $C_{i,j}$ as well as the \textit{distance to interaction} $D_{i,j}$.
 }
 \label{fig:binaryencounter}
\end{figure}
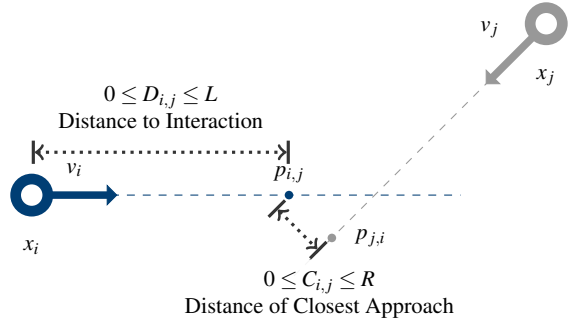

Further useful quantities can be derived from $\tau_{i,j}$; see \cref{fig:binaryencounter}. The \textit{point of closest approach of $i$ to $j$}, $p_{i,j}$, is the point along the trajectory of $i$ where the agents will be closest:
\begin{equation}
 \norm{p_{i,j}-p_{j,i}}=\min_{t\in\mathbb{R}}\norm{x_{i}(t)-x_{j}(t)}
 \quad\text{and}\quad
 p_{i,j}=x_i+v_i\tau_{i,j}.
\end{equation}
The \textit{distance to interaction of $i$ with $j$}, $D_{i,j}$, is the distance of $i$ to $p_{i,j}$, that is:
\begin{equation}\label{eq:Dij}
 D_{i,j}=\norm{p_{i,j}-x_{i}}=\tau_{i,j}\norm{v_i}
 =-\frac{\prt{x_j-x_i}\cdot\prt{v_j-v_i}}{\norm{v_j-v_i}^2}\norm{v_i}.
\end{equation}
Last but not least, the \textit{distance of closest approach of $i$ and $j$}, $C_{i,j}$:
\begin{equation}\label{eq:Cij}
 C_{i,j}=\norm{p_{i,j}-p_{j,i}}
 =\prt{\norm{x_j-x_i}^2-\frac{\prt{\prt{x_j-x_i}\cdot\prt{v_j-v_i}}^2}{\norm{v_j-v_i}^2}}^{\frac{1}{2}}.
\end{equation}
It is worth noting that $\tau_{i,j}$ and $C_{i,j}$ are symmetric for $i$ and $j$.

\subsubsection{Assumptions on the Heuristics}\label{sec:assumptions}

We will make a number of assumptions about the quantities derived above which will dictate what encounters can be considered by pedestrians:
\begin{enumerate}
 \item $\tau_{i,j}>0$ and $D_{i,j}>0$. From its definition on \eqref{eq:tij}, it is clear that $\tau_{i,j}$ will be a negative number whenever $\prt{x_j-x_i}\cdot\prt{v_j-v_i}>0$. Geometrically, the inner product is positive whenever pedestrians $i$ and $j$ are moving away from each other; see \cref{fig:approachingpedestrians}. As such, we can limit our consideration to positive values of $\tau_{i,j}$ (and by extension $D_{i,j}$), since pedestrians that are separating will simply ignore each other because there is no potential collision.
       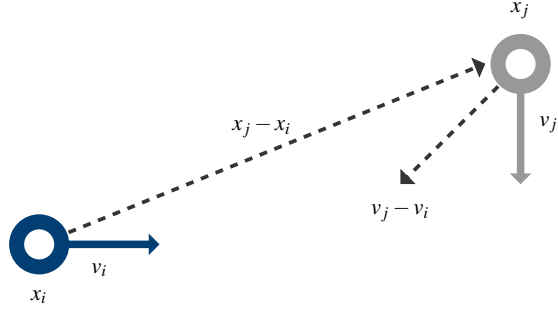
\begin{figure}
        \sidecaption
        \centering
        \begin{tikzpicture}[scale=0.80]
 \def\xi{0}
 \def\yi{0}
 \def\ui{2.0}
 \def\vi{0}
 \def\xj{8}
 \def\yj{3}
 \def\uj{0}
 \def\vj{-2.0}

 \path[thicker line small arrows m=0.5mm in color3, dashed] (\xi,\yi) -- ({\xj-0.6},\yj) node[black,midway,above=0.1cm]{$x_j-x_i$};
 \path[thicker line small arrows m=1mm in color1] (\xi,\yi) -- ({\xi+\ui},{\yi+\vi}) node[black,midway,below=0.1cm]{$v_i$};
 \path[thicker line small arrows m=1mm in color2] (\xj,\yj) -- ({\xj+\uj},{\yj+\vj}) node[black,midway,right=0.1cm]{$v_j$};
 \path[thicker line small arrows m=0.5mm in color3, dashed] (\xj,\yj) -- ({\xj+\uj-\ui},{\yj+\vj-\vi}) node[black,below=0.1cm]{$v_j-v_i$};

 \fill[color1] (\xi,\yi) circle (0.50cm);
 \fill[white] (\xi,\yi) circle (0.25cm) node[black, below=0.5cm]{$x_i$};
 \fill[color2] (\xj,\yj) circle (0.50cm);
 \fill[white] (\xj,\yj) circle (0.25cm) node[black, above=0.5cm]{$x_j$};
\end{tikzpicture}
        \caption{Approaching pedestrians. Agent $i$ only considers $j$ if they are approaching, i.e. if $\prt{x_j-x_i}\cdot\prt{v_j-v_i}<0$.}
        \label{fig:approachingpedestrians}
       \end{figure}
 \item $D_{i,j}<L$. The bound on the distance to interaction reflects the fact that $i$ does not react to obstacles beyond a certain distance. $L$ can be thought of as a \textit{visual horizon} for pedestrians that limits their interactions. The encounters will be ignored unless they are sufficiently close.
 \item $C_{i,j}<R$. In the same vein, the bound on the distance of closest approach points to the fact that $i$ does not account for obstacles that will never be close by. $R$ is a measure of the \textit{personal space} of the agents. Unless this space is invaded, there is no reaction.
 \item Only visible pedestrians are considered. An agent $i$ cannot consider $j$ for collision avoidance without seeing them since all the heuristics are derived from optical stimuli. Pedestrian $j$ is visible from $i$'s point of view whenever:
       \begin{equation}
        \frac{\prt{x_j-x_i}\cdot v_i}{\norm{x_j-x_i}\norm{v_i}}>\cos(\vartheta/2),
       \end{equation}
       for agents with a \textit{horizontal field of view} $\vartheta$. In humans, $\vartheta=7\pi/6$ \cite{Traquair1876}.
\end{enumerate}

\subsubsection{Global Encounters}

While the pairwise encounters of \cref{sec:pairwiseencounters} are the typical interaction between pedestrians in situations of low agent concentration, more complex configurations are expected in the higher density regimes. A characterisation of arrangements involving more than two pedestrians is required.

The pairwise heuristics presented above can be combined to render \textit{global heuristics} describing more intricate encounters. In order to do so, we will assume that pedestrians react first to whichever interaction is closest. Given two potential collisions, the agent will avoid the nearer one first, and then deal with the further one if necessary.

Bearing in mind the order of interactions, it is straightforward to define the \textit{global distance to interaction for $i$, $D_i$}:
\begin{equation}
 D_i=\min_{j} \set{D_{i,j}}
 \quad\text{for suitable }j.
 \label{eq:Di}
\end{equation}
Admissible agents $j$ for the minimisation are the \textit{agents perceived by $i$}, i.e. those satisfying the assumptions from \cref{sec:assumptions} together with $i$. The definition of the \textit{global distance of closest approach for $i$, $C_i$} follows immediately as a consequence of this choice:
\begin{equation}
 C_i=C_{i,j}\quad\text{for $j$ that minimises \eqref{eq:Di}}.
 \label{eq:Ci}
\end{equation}
Both of these global heuristics will be used to inform the pedestrian during his choice of direction on the next phase.

\subsection{Decision Stage}
The decision stage is the second and last step towards collision avoidance in the model. Following the obtention of global heuristics during the previous phase, pedestrians must now employ said heuristics in order to decide how to alter their trajectory.

In deciding on a new path, it is helpful to portray the heuristics of each pedestrian as functions of their velocity. Each of the quantities is transformed into a map by allowing $v_i$ to become a variable: $D_{i,j}\prt{v}$ and $C_{i,j}\prt{v}$. In \cite{Degond2013} pedestrians are assumed to have a uniform speed $s$, and thus the heuristics are purely functions of the \textit{direction} $\omega$.

Each agent can use the perceived overall heuristics to inform their choice of velocity. The decision obeys two antagonist interests: navigation towards a target and obstacle avoidance. Pedestrians will maintain their \textit{target velocity} $v_i^*$ consisting of their \textit{comfort speed} $s_i^*$ and their \textit{target direction} $\omega_i^*$ whenever possible. They will only deviate from this velocity when a collision is about to occur. Agents will then consider directions within their field of view and make a choice that resolves the collisions while deviating minimally from the target.

\begin{figure}
 \sidecaption
 \centering
 \begin{tikzpicture}[scale=0.6]
 \def\l{2.0}
 \def\ll{2.5}
 \def\L{6.0}

 \draw [dashed, color1] ({\ll*cos(15)},{\ll*sin(15)}) -- ({\L*cos(15)},{\L*sin(15)});
 \draw [dashed, color1] ({\ll*cos(5)},{\ll*sin(5)}) -- ({\L*cos(5)},{\L*sin(5)});
 \draw [dashed, color1] ({\ll*cos(-5)},{\ll*sin(-5)}) -- ({\L*cos(-5)},{\L*sin(-5)});
 \draw [dashed, color1] ({\ll*cos(-15)},{\ll*sin(-15)}) -- ({\L*cos(-15)},{\L*sin(-15)});
 \draw [dashed, color1] (0,0) -- ({\L*cos(10)},{\L*sin(10)});
 \draw [dashed, color1] (0,0) -- ({\L*cos(-10)},{\L*sin(-10)});
 \draw [dashed, color1] (0,0) -- ({\L*cos(20)},{\L*sin(20)});
 \draw [dashed, color1] (0,0) -- ({\L*cos(-20)},{\L*sin(-20)});
 \draw [very thick, dashed, color1] (0,0) -- (\L,0);

 \path[thicker line small arrows m=1mm in color1] (0,0) -- (2.0,0) ;

 \fill[color1] (0,0) circle (0.50cm);
 \fill[white] (0,0) circle (0.25cm);

 \draw [|<->|, very thick, dotted, color3, domain=20:-20] plot ({1.10*\L*cos(\x)}, {1.10*\L*sin(\x)});

 \def\xj{-1}
 \def\yj{2.5}
 \def\tj{-10}

 \def\xk{3}
 \def\yk{-2.5}
 \def\tk{35}

 \def\xl{9}
 \def\yl{1.75}
 \def\tl{195}

 \def\xm{10}
 \def\ym{-1.5}
 \def\tm{135}

 \path[thicker line small arrows m=1mm in gray] (\xj,\yj) -- ({\xj+\l*cos(\tj)},{\yj+\l*sin(\tj)});
 \fill[gray] (\xj,\yj) circle (0.50cm);
 \fill[white] (\xj,\yj) circle (0.25cm);

 \path[thicker line small arrows m=1mm in gray] (\xk,\yk) -- ({\xk+\l*cos(\tk)},{\yk+\l*sin(\tk)});
 \fill[gray] (\xk,\yk) circle (0.50cm);
 \fill[white] (\xk,\yk) circle (0.25cm);

 \path[thicker line small arrows m=1mm in gray] (\xl,\yl) -- ({\xl+\l*cos(\tl)},{\yl+\l*sin(\tl)});
 \fill[gray] (\xl,\yl) circle (0.50cm);
 \fill[white] (\xl,\yl) circle (0.25cm);

 \path[thicker line small arrows m=1mm in gray] (\xm,\ym) -- ({\xm+\l*cos(\tm)},{\ym+\l*sin(\tm)});
 \fill[gray] (\xm,\ym) circle (0.50cm);
 \fill[white] (\xm,\ym) circle (0.25cm);

 \node[black,draw=none,right=0.2cm,opacity=.9,text opacity=1] at ({1.10*\L*cos(-20)}, {1.10*\L*sin(-20)}) {Possible paths};
\end{tikzpicture}
 \caption{A pedestrian considers possible paths. They intend to remain in motion towards the target but must avoid colliding with other agents. Both of these goals must be considered in deciding on a new direction.}
 \label{fig:possiblepaths}
\end{figure}
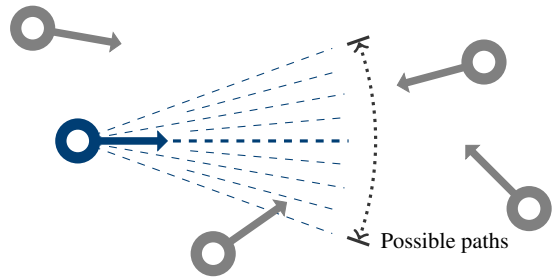

\subsubsection{The Decision Potential}

The task of finding a suitable direction can be cast as a game-theoretical problem involving the minimisation of a cost. We formulate this task through the \textit{decision potential} or \textit{decision function} $\Phi_i(v)$. This function must reflect the wishes and tendencies of the agents: to move according to a navigation goal and to resolve potential collisions. Suitable velocities, which are clear of interactions and oriented towards the target, will yield low values of $\Phi_i$. Less suitable velocities, either incurring collisions or leading away from the target, will have higher costs.

Each pedestrian will choose the optimal velocity according to the decision potential and evolve through an interval of time $\Delta t$ before making a new choice. This minimisation reflects the decision-making of pedestrians, who strive to always move towards their target while avoiding collisions with a minimal amount of steering.

Letting $u_i^n$ be the minimiser at the $n$-th step, the evolution of each agent will be written as a difference equation:
\begin{align}
 x^{n+1}_i & =x^{n}_i+u^{n}_i\Delta t
 ,         &
 u^{n}_i   & = \argmin_{v}\Phi^n_i\prt{v}
 .
 \label{eq:modelopti}
\end{align}

\subsubsection{A Choice of Potential}\label{sec:choiceofphi}
The formulation of \cite{Degond2013} proposes the decision function:
\begin{equation}
 \Phi_{i}\prt{v}=\frac{k}{2}\norm{D_iv-Lv_i^*}^2,\label{eq:phi}
\end{equation}
where $k$ is a positive constant and $v_i^*$ is the \textit{target velocity} of agent $i$.

The workings of each individual tendency are reflected on the potential. $\Phi$ penalises deviations of $v$ from $v_i^*$ (moving away from the target) as well as deviations of $D_i$ from $L$ (potential collisions). In the absence of other pedestrians $D_i\equiv L$ and thus $\Phi_i\prt{v}\equiv kL^2\norm{v-v_i^*}^2/2$, a convex cost with unique minimum $v=v_i^*$.

\subsection{A Gradient-Based Formulation} The model as presented thus far is computationally costly; numerically solving  two minimisation problems per pedestrian per step is not practical. Furthermore the choice of time step can be problematic: $\Delta t< {D_i}/{\norm{v_i}}$ is required in order to successfully resolve collisions, but too small a step can result on the velocity of an agent varying erratically when interacting with a large number of pedestrians in a dense setting.

An alternative is to formulate the decision-making process in terms of the gradient of the decision function through the differential equation:
\begin{align}\label{eq:modelgradient}
 \der{x_i}{t} & =v_i, & \der{v_i}{t}=-\nabla_{v}\Phi_i\prt{v_i}.
\end{align}

While \eqref{eq:modelgradient} is governed by similar principles to those of the optimisation model, it yields a continuous dynamic that accounts for a decision-making process similar to that of \eqref{eq:modelopti} but lets agents gradually shift towards suitable velocities by descending the gradient of $\Phi$.

\subsubsection{Optimality versus Efficiency}

The gradient scheme \eqref{eq:modelgradient} provides an efficient alternative to the optimisation formulation, as the numerical computation of $\nabla \Phi$ once per agent per time step is inexpensive when compared with the performance of any optimisation scheme that may be used to approximate the global minimiser.

The optimisation scheme \eqref{eq:modelopti} remains the preferable method for simulations involving a small number of agents and low densities, as it guarantees the optimal choice of velocity for all agents. Gradient descent may simply not reach the global minimum that would be found otherwise, as it may stall upon reaching a local minimum. However, in high-density regimes the decision potential becomes volatile, as the increased interaction rate rapidly changes the cost of each velocity; local minima are unlikely to persist. Furthermore, the global minimum may change abruptly and repeatedly, which would result in rapid switching of agent directions. A gradient method appears more suitable, as simply shifting away from high-cost velocities might be sufficient for collision avoidance.

\subsection{Summary of the General Model}
Consider $N$ pedestrians, where agent $i$ has position $x_i$, velocity $v_i$, and target velocity $v_i^*$. The dynamics will be given by the solution to either \eqref{eq:modelopti}:
\begin{align}
 x^{n+1}_i & =x^{n}_i+u^{n}_i\Delta t,
           & u^{n}_i                   & =\argmin_{v}\Phi^n_i\prt{v},
\end{align}
or \eqref{eq:modelgradient}, namely:
\begin{align}
 \der{x_i}{t} & =v_i, & \der{v_i}{t}=-\nabla_{v}\Phi_i\prt{v_i}.
\end{align}
The evaluation of the decision potential $\Phi$ is as follows:
\begin{enumerate}
 \item For each pair of agents $i$ and $j$, compute the heuristics $D_{i,j}$ and $C_{i,j}$ as defined in \eqref{eq:Dij} and \eqref{eq:Cij}:
       \begin{align}
        D_{i,j}= & -\frac{\prt{x_j-x_i}\cdot\prt{v_j-v_i}}{\norm{v_j-v_i}^2}\norm{v_i},
        \\C_{i,j}=&\prt{\norm{x_j-x_i}^2-\frac{\prt{\prt{x_j-x_i}\cdot\prt{v_j-v_i}}^2}{\norm{v_j-v_i}^2}}^{\frac{1}{2}}.
       \end{align}
 \item Decide whether $i$ will take $j$ into account using the conditions from \cref{sec:assumptions}:
       \begin{align}
        D_{i,j} & <L,
        \\C_{i,j}&<R,
        \\\prt{x_j-x_i}\cdot\prt{v_j-v_i}&<0,
        \\\cos(\vartheta/2)&<\frac{\prt{x_j-x_i}\cdot v_i}{\norm{x_j-x_i}\norm{v_i}}.
       \end{align}
 \item Obtain overall heuristics $D_{i}$ and $C_{i}$ as defined in \eqref{eq:Di} and \eqref{eq:Ci}:
       \begin{align}
        D_i & =D_{i,j^*},
            & C_i         & =C_{i,j^*},
            & j^*         & =\argmin_{j}\set{D_{i,j}}.
       \end{align}
 \item Use the global heuristics to construct the cost function $\Phi_i$ as defined in \eqref{eq:phi}.
\end{enumerate}

\section{Towards a High Density Model}\label{sec:highdensity}
The model of \cite{Degond2013} is explicitly formulated for low pedestrian densities, where all agents have a uniform, constant speed. Moving at such a speed is simply infeasible in more congested regimes, as agents will be required to slow down or even stop completely in order to avoid collisions. Even groups where all the agents move in the same direction will show a reduction on speed whenever the intra-crowd density is beyond some thresholds \cite{Weidmann1993}.

This section presents a study of the original formulation, followed by a series of variations of the model that aim to extend its validity to pedestrians with variable speeds and scenarios of higher densities.

\subsection{A Frontal Collision}\label{sec:frontalcollision}
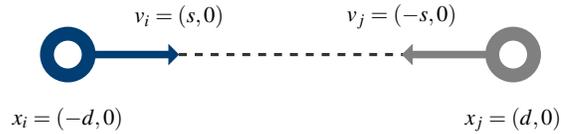
\begin{figure}
 \sidecaption
 \centering
 \begin{tikzpicture}[scale=0.74]
 \def\l{2.0}
 \def\ll{2.5}
 \def\L{6.0}

 \def\xj{8}
 \def\yj{0}
 \def\tj{180}

 \draw [very thick, dashed, color3] (0,0) -- (\xj,\yj);

 \path[thicker line small arrows m=1mm in color1] (0,0) -- (2.0,0)  node[black, above=0.20cm]{$v_i=\prt{s,0}$};
 \fill[color1] (0,0) circle (0.50cm) node[black, below=0.60cm]{$x_i=\prt{-d,0}$};
 \fill[white] (0,0) circle (0.25cm);

 \path[thicker line small arrows m=1mm in gray] (\xj,\yj) -- ({\xj+\l*cos(\tj)},{\yj+\l*sin(\tj)}) node[black, above=0.20cm]{$v_j=\prt{-s,0}$};
 \fill[gray] (\xj,\yj) circle (0.50cm) node[black, below=0.60cm]{$x_j=\prt{d,0}$};
 \fill[white] (\xj,\yj) circle (0.25cm);
\end{tikzpicture}
 \caption{
  Study of a frontal collision. Agents $i$ and $j$ are placed in a simple arrangement, at a distance $2d$ and approaching in a straight line, each with speed $s$. The possible velocities for $i$ are parametrised as $v=s\prt{\cos\theta,\sin\theta}$.
 }
 \label{fig:frontalcollision}
\end{figure}

A natural range of validity for the model can be obtained by considering a frontal encounter of two pedestrians and studying the corresponding decision function under the framework of \eqref{eq:modelopti}. For simplicity, we choose:
\begin{align}
 x_i & =\prt{-d,0}, & x_j & =\prt{d,0},  \\\nonumber
 v_i & =\prt{s,0},  & v_j & =\prt{-s,0},
\end{align}
for given distance $d$ and speed $s$. Considering a test velocity for agent $i$, $v=s\prt{\cos\theta,\sin\theta}$, and ignoring variations on the speed as per the original formulation, it can be readily verified:
\begin{align}
 \tau_{i,j}\prt{v} & =\frac{d}{s}, & D_{i,j}\prt{v} & =d, & C_{i,j}\prt{v} & =\sqrt{2d^2\prt{1-\cos\theta}}.
\end{align}

\begin{figure}
 \centering
 \sidecaption
 \includegraphics[width=0.65\textwidth]{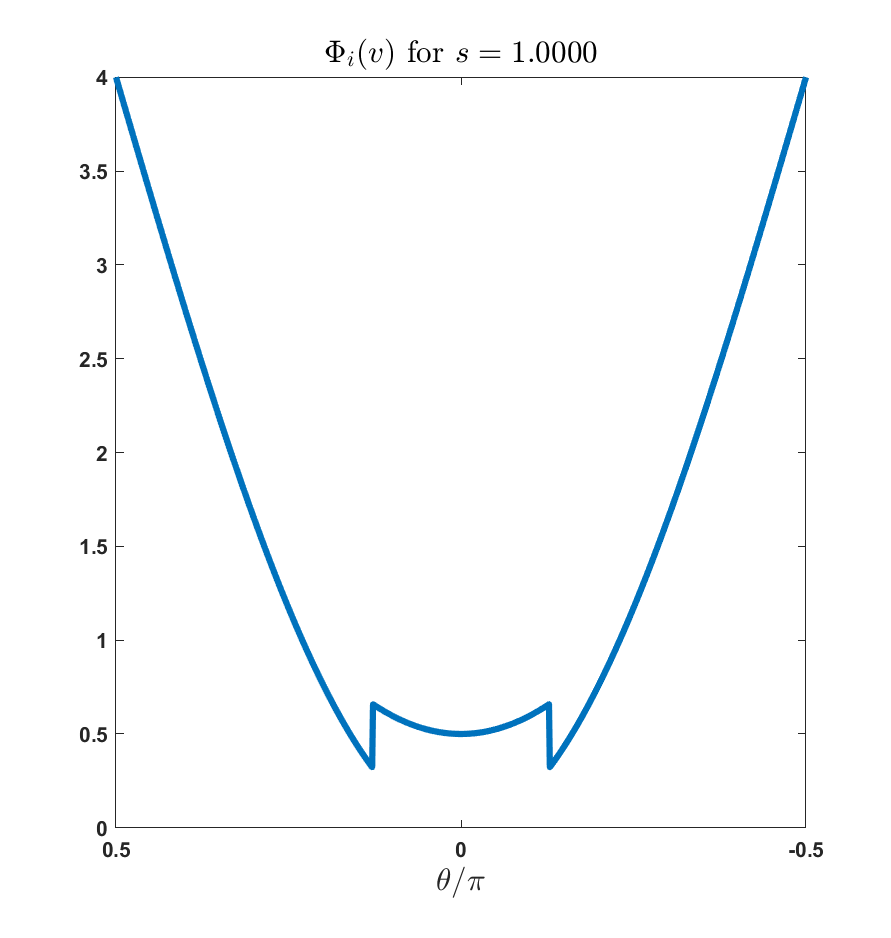}
 \includegraphics[width=0.65\textwidth]{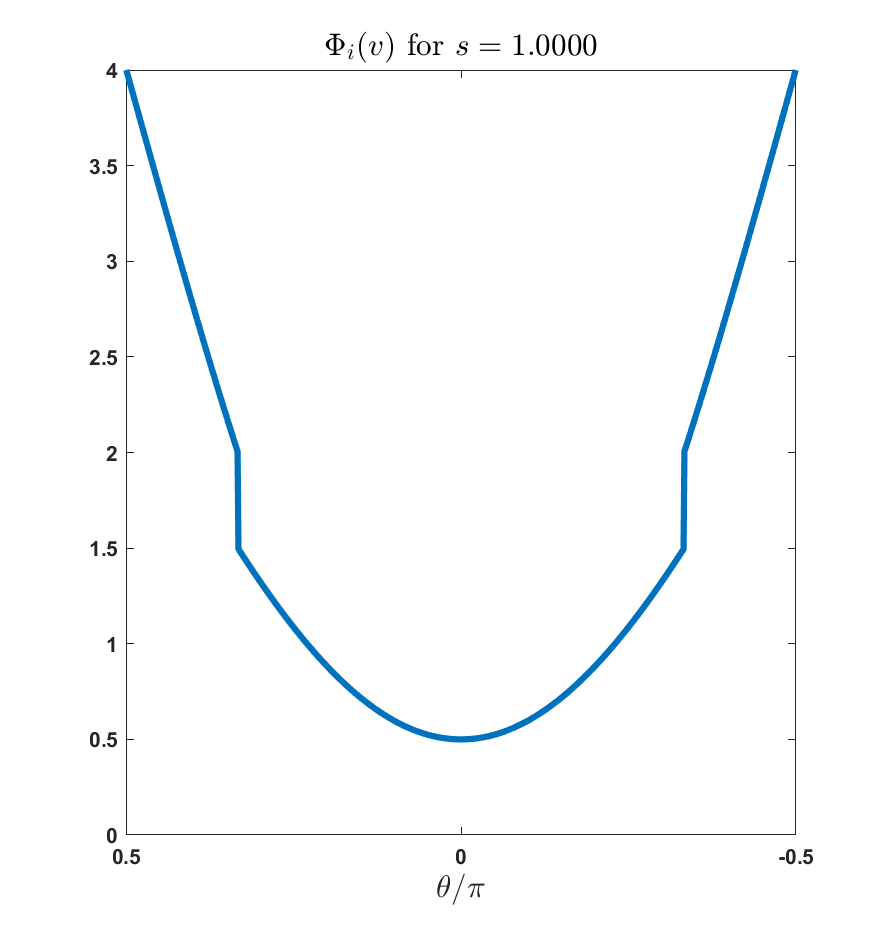}
 \caption{
  Comparison of $\Phi_i$ for the frontal collision. $d=1.0, s=1.0, k=1.0$.
  \textbf{Top:} $L=2.0, R=0.4$. The global minima correctly fall to either side of the region of interaction.
  \textbf{Bottom:} $L=2.0, R=1.0$. The global minimum appears at $\theta=0$, leading to a collision.
 }
 \label{fig:comparisonphifrontalcollision}
\end{figure}

$D_i(v)$ is thus equal to $d$ if $i$ perceives $j$ and equal to $L$ otherwise. The decision function reduces to:
\begin{equation}
 \Phi_i\prt{v}=
 \begin{dcases}
  ks^2\prt{
   \frac{L^2+d^2}{2}
   -Ld\cos\theta
  }
   & \text{if $i$ perceives $j$,}
  \\
  ks^2L^2\prt{1-\cos\theta}
   & \text{otherwise.}
 \end{dcases}
\end{equation}
If we assume $d<L$, then the perception of $j$ by $i$ is conditioned solely by $C_i<R$, which yields:
\begin{equation}\label{eq:perceptionAngle}
 \abs{\theta}<\theta^*\defeq \arccos\prt{1-\frac{R^2}{2d^2}}.
\end{equation}

Each segment of $\Phi$ now has clear minima: the perception section,
where $\theta$ verifies \eqref{eq:perceptionAngle}, has its minimum at $\theta=0$ and takes the value $\Phi\prt{0}={ks^2}\prt{L-d}^2/{2}$; the remaining section has its minima at the boundary of the two regions,
$\abs{\theta}=\theta^*$,
where $\Phi\prt{\theta^*}={ks^2}{L^2R^2}/{2d^2}$.

In order for the collision to be successfully resolved, the central minimum must not be selected in \eqref{eq:modelopti}. The wrong choice is made whenever $\Phi\prt{0}<\Phi\prt{\theta^*}$, and this criterion can be used to ascertain validity:
\begin{align}
 \Phi\prt{0}<\Phi\prt{\theta^*}\qquad\iff\qquad d^2\prt{L-d}^2<{L^2R^2},\label{eq:inequalityPolynomial}
\end{align}
since $k>0, s>0$.

Given that $L$ and $R$ are constants, the resulting polynomial of $d$, $p\prt{d}\defeq d^2\prt{L-d}^2$, encodes the fate of the interaction.
In order to navigate the collision successfully, the condition of \eqref{eq:inequalityPolynomial}, rewritten as $p(d)<L^2R^2$, must be violated at some point as the agents $i$ and $j$ approach. Recall $D_{i,j}=d$; if the condition is met for large $d$ (say, $d=L$) and continues to hold as $d\rightarrow 0$, the agents will collide.

The polynomial $p\prt{d}$ is a monic quartic with double roots at $d=0$ and $d=L$. Naturally, this suggests a single maximum at $d=L/2$, with $p\prt{L/2}=L^4/2^4$. This extremum must exceed $L^2R^2$; otherwise \eqref{eq:inequalityPolynomial} is valid for all $d\in\prt{0,L}$, and a collision occurs. Simplifying:
\begin{align}
 p\prt{\frac{L}{2}}<L^2R^2\qquad\iff\qquad L<4R.\label{eq:inequalityNaturalScale}
\end{align}

The comparison yields  $L<4R$ as a sufficient condition for the model to fail by means of the interaction minimum ($\theta=0$) becoming the global minimum, resulting in a collision (see \cref{fig:comparisonphifrontalcollision}). \eqref{eq:inequalityNaturalScale} amounts to a natural scale of the model, which is only valid in regimes where $L\gg R$, i.e. the visual horizon is much larger than the personal space of the agents. A large horizon is a characteristic of low-density scenarios, which justifies the original choice of prescribing the model for such regimes and constant speeds. Denser scenarios typically show $L\sim R$, as agents can only consider nearby interactions, which takes the model outside its region of validity.

\subsection{Grading by Collision Severity}

\begin{figure}
 \centering
 \sidecaption
 \includegraphics[width=0.65\textwidth]{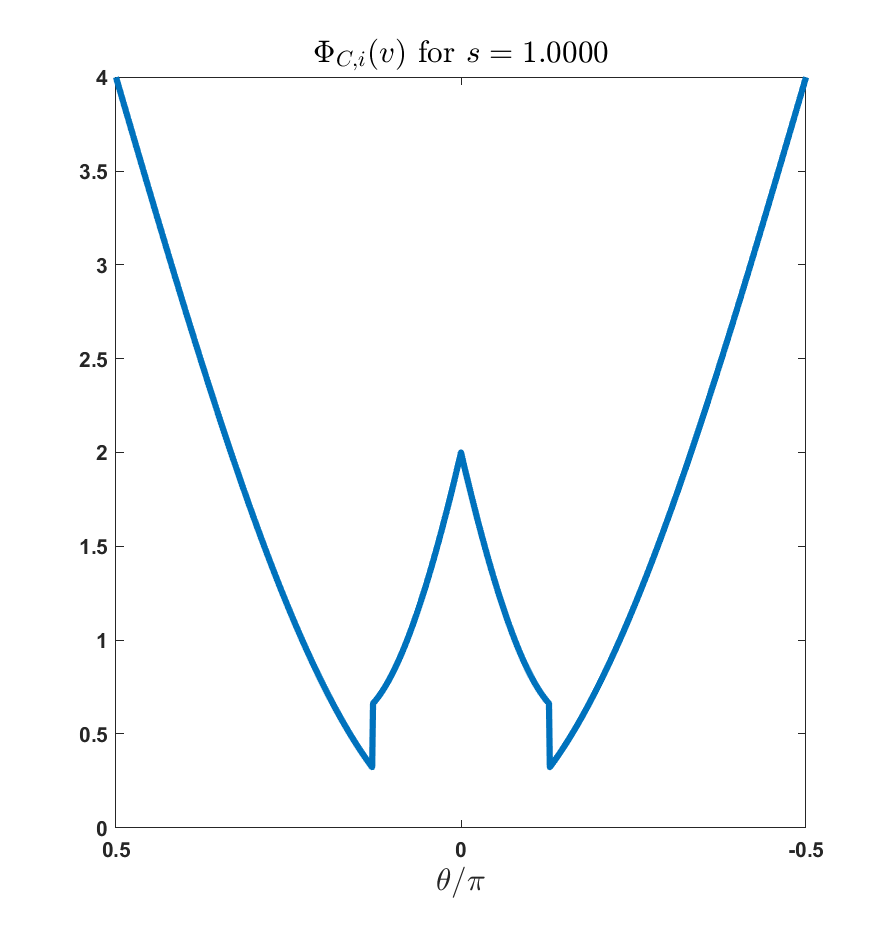}
 \includegraphics[width=0.65\textwidth]{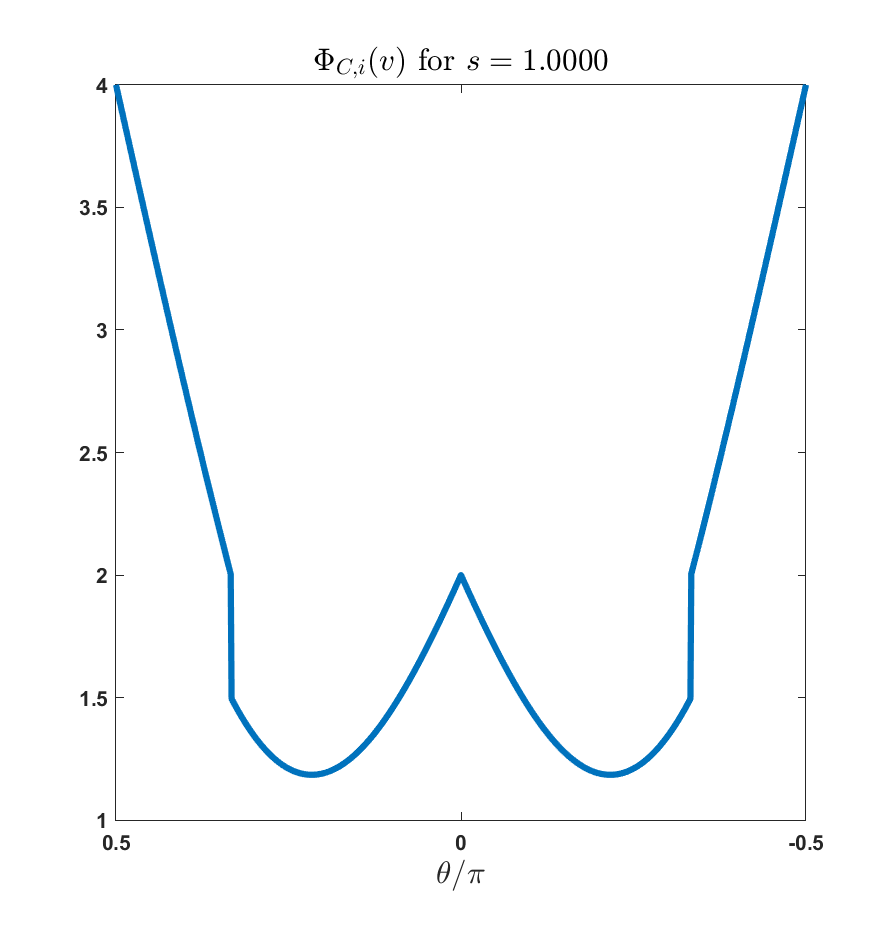}
 \caption{
  Comparison of $\Phi_{C,i}$ for the frontal collision. $d=1.0, s=1.0, k=1.0$. Contrary to \cref{fig:comparisonphifrontalcollision}, both sets of parameters yield correct shapes. In particular, the central minima have disappeared.
  \textbf{Top:} $L=2.0, R=0.4$.
  \textbf{Bottom:} $L=2.0, R=1.0$.
 }
 \label{fig:comparisonphiCfrontalcollision}
\end{figure}

Following the analysis of the original model, and considering the natural scale found in \cref{sec:frontalcollision}, we aim to extend this model to higher-density regimes where the criterion $L\leq 4R$ might be met.

A straightforward solution to the model choosing the erroneous central minimum is simply removing the minimum altogether. Notice that, while both global heuristics $D_i$ and $C_i$ are used to discern what pedestrians need to be considered by $i$ for collision avoidance, only $D_i$ appears explicitly on the potential. Recall that $C_i$ is a measure of the severity of an interaction, ranging from $0$ for a full collision to $R$ for no interaction at all. Thus penalising deviations of $C_i$ from $R$ on top of the existing penalisations in \eqref{eq:phi} will result in a higher cost at points of full collision, namely, the troubling point $\theta=0$. We propose:
\begin{equation}
 \Phi_{C,i}\prt{v}=\frac{k}{2R^2}\norm{D_iC_iv-LRv_i^*}^2.\label{eq:phiC}
\end{equation}

\begin{figure}
 \centering
 \sidecaption
 \begin{tabular}{ c c c }
  $t=0.0$ &                & $t=4.1$
  \\ &&  \\
  \begin{tikzpicture}[scale=0.5]
   \fill[color4] (-5, -2) circle(0.05 cm);
\fill[color4] (-5, -1) circle(0.05 cm);
\fill[color4] (-5, 0) circle(0.05 cm);
\fill[color4] (-5, 1) circle(0.05 cm);
\fill[color4] (-5, 2) circle(0.05 cm);
\fill[color4] (-4, -2) circle(0.05 cm);
\fill[color4] (-4, -1) circle(0.05 cm);
\fill[color4] (-4, 0) circle(0.05 cm);
\fill[color4] (-4, 1) circle(0.05 cm);
\fill[color4] (-4, 2) circle(0.05 cm);
\fill[color4] (-3, -2) circle(0.05 cm);
\fill[color4] (-3, -1) circle(0.05 cm);
\fill[color4] (-3, 0) circle(0.05 cm);
\fill[color4] (-3, 1) circle(0.05 cm);
\fill[color4] (-3, 2) circle(0.05 cm);
\fill[color4] (-2, -2) circle(0.05 cm);
\fill[color4] (-2, -1) circle(0.05 cm);
\fill[color4] (-2, 0) circle(0.05 cm);
\fill[color4] (-2, 1) circle(0.05 cm);
\fill[color4] (-2, 2) circle(0.05 cm);
\fill[color4] (-1, -2) circle(0.05 cm);
\fill[color4] (-1, -1) circle(0.05 cm);
\fill[color4] (-1, 0) circle(0.05 cm);
\fill[color4] (-1, 1) circle(0.05 cm);
\fill[color4] (-1, 2) circle(0.05 cm);
\fill[color4] (0, -2) circle(0.05 cm);
\fill[color4] (0, -1) circle(0.05 cm);
\fill[color4] (0, 0) circle(0.05 cm);
\fill[color4] (0, 1) circle(0.05 cm);
\fill[color4] (0, 2) circle(0.05 cm);
\fill[color4] (1, -2) circle(0.05 cm);
\fill[color4] (1, -1) circle(0.05 cm);
\fill[color4] (1, 0) circle(0.05 cm);
\fill[color4] (1, 1) circle(0.05 cm);
\fill[color4] (1, 2) circle(0.05 cm);
\fill[color4] (2, -2) circle(0.05 cm);
\fill[color4] (2, -1) circle(0.05 cm);
\fill[color4] (2, 0) circle(0.05 cm);
\fill[color4] (2, 1) circle(0.05 cm);
\fill[color4] (2, 2) circle(0.05 cm);
\fill[color4] (3, -2) circle(0.05 cm);
\fill[color4] (3, -1) circle(0.05 cm);
\fill[color4] (3, 0) circle(0.05 cm);
\fill[color4] (3, 1) circle(0.05 cm);
\fill[color4] (3, 2) circle(0.05 cm);
\fill[color4] (4, -2) circle(0.05 cm);
\fill[color4] (4, -1) circle(0.05 cm);
\fill[color4] (4, 0) circle(0.05 cm);
\fill[color4] (4, 1) circle(0.05 cm);
\fill[color4] (4, 2) circle(0.05 cm);
\fill[color4] (5, -2) circle(0.05 cm);
\fill[color4] (5, -1) circle(0.05 cm);
\fill[color4] (5, 0) circle(0.05 cm);
\fill[color4] (5, 1) circle(0.05 cm);
\fill[color4] (5, 2) circle(0.05 cm);
\path[thicker line small arrows m=1mm in color2] (-4, 0) -- (-2.5, 0);
\fill[color2] (-4, 0) circle(0.35 cm);
\fill[white] (-4, 0) circle(0.175 cm);
\path[thicker line small arrows m=1mm in color2] (4, 0.004) -- (2.5, 0.004);
\fill[color2] (4, 0.004) circle(0.35 cm);
\fill[white] (4, 0.004) circle(0.175 cm);
  \end{tikzpicture}
          & \hspace{0.5cm} &
  \begin{tikzpicture}[scale=0.5]
   \fill[color4] (-5, -2) circle(0.05 cm);
\fill[color4] (-5, -1) circle(0.05 cm);
\fill[color4] (-5, 0) circle(0.05 cm);
\fill[color4] (-5, 1) circle(0.05 cm);
\fill[color4] (-5, 2) circle(0.05 cm);
\fill[color4] (-4, -2) circle(0.05 cm);
\fill[color4] (-4, -1) circle(0.05 cm);
\fill[color4] (-4, 0) circle(0.05 cm);
\fill[color4] (-4, 1) circle(0.05 cm);
\fill[color4] (-4, 2) circle(0.05 cm);
\fill[color4] (-3, -2) circle(0.05 cm);
\fill[color4] (-3, -1) circle(0.05 cm);
\fill[color4] (-3, 0) circle(0.05 cm);
\fill[color4] (-3, 1) circle(0.05 cm);
\fill[color4] (-3, 2) circle(0.05 cm);
\fill[color4] (-2, -2) circle(0.05 cm);
\fill[color4] (-2, -1) circle(0.05 cm);
\fill[color4] (-2, 0) circle(0.05 cm);
\fill[color4] (-2, 1) circle(0.05 cm);
\fill[color4] (-2, 2) circle(0.05 cm);
\fill[color4] (-1, -2) circle(0.05 cm);
\fill[color4] (-1, -1) circle(0.05 cm);
\fill[color4] (-1, 0) circle(0.05 cm);
\fill[color4] (-1, 1) circle(0.05 cm);
\fill[color4] (-1, 2) circle(0.05 cm);
\fill[color4] (0, -2) circle(0.05 cm);
\fill[color4] (0, -1) circle(0.05 cm);
\fill[color4] (0, 0) circle(0.05 cm);
\fill[color4] (0, 1) circle(0.05 cm);
\fill[color4] (0, 2) circle(0.05 cm);
\fill[color4] (1, -2) circle(0.05 cm);
\fill[color4] (1, -1) circle(0.05 cm);
\fill[color4] (1, 0) circle(0.05 cm);
\fill[color4] (1, 1) circle(0.05 cm);
\fill[color4] (1, 2) circle(0.05 cm);
\fill[color4] (2, -2) circle(0.05 cm);
\fill[color4] (2, -1) circle(0.05 cm);
\fill[color4] (2, 0) circle(0.05 cm);
\fill[color4] (2, 1) circle(0.05 cm);
\fill[color4] (2, 2) circle(0.05 cm);
\fill[color4] (3, -2) circle(0.05 cm);
\fill[color4] (3, -1) circle(0.05 cm);
\fill[color4] (3, 0) circle(0.05 cm);
\fill[color4] (3, 1) circle(0.05 cm);
\fill[color4] (3, 2) circle(0.05 cm);
\fill[color4] (4, -2) circle(0.05 cm);
\fill[color4] (4, -1) circle(0.05 cm);
\fill[color4] (4, 0) circle(0.05 cm);
\fill[color4] (4, 1) circle(0.05 cm);
\fill[color4] (4, 2) circle(0.05 cm);
\fill[color4] (5, -2) circle(0.05 cm);
\fill[color4] (5, -1) circle(0.05 cm);
\fill[color4] (5, 0) circle(0.05 cm);
\fill[color4] (5, 1) circle(0.05 cm);
\fill[color4] (5, 2) circle(0.05 cm);
\path[thicker line small arrows m=1mm in color2] (0.22039195215585403, -0.5292187561094341) -- (1.7233445708249024, -0.6623041515215065);
\fill[color2] (0.22039195215585403, -0.5292187561094341) circle(0.35 cm);
\fill[white] (0.22039195215585403, -0.5292187561094341) circle(0.175 cm);
\path[thicker line small arrows m=1mm in color2] (-0.22039195215585403, 0.5332187561094341) -- (-1.7233445708249024, 0.6663041515215065);
\fill[color2] (-0.22039195215585403, 0.5332187561094341) circle(0.35 cm);
\fill[white] (-0.22039195215585403, 0.5332187561094341) circle(0.175 cm);
  \end{tikzpicture}
  \\  \vspace{0.5cm} &&  \\
  $t=1.9$ &                & $t=4.6$
  \\ &&  \\
  \begin{tikzpicture}[scale=0.5]
   \fill[color4] (-5, -2) circle(0.05 cm);
\fill[color4] (-5, -1) circle(0.05 cm);
\fill[color4] (-5, 0) circle(0.05 cm);
\fill[color4] (-5, 1) circle(0.05 cm);
\fill[color4] (-5, 2) circle(0.05 cm);
\fill[color4] (-4, -2) circle(0.05 cm);
\fill[color4] (-4, -1) circle(0.05 cm);
\fill[color4] (-4, 0) circle(0.05 cm);
\fill[color4] (-4, 1) circle(0.05 cm);
\fill[color4] (-4, 2) circle(0.05 cm);
\fill[color4] (-3, -2) circle(0.05 cm);
\fill[color4] (-3, -1) circle(0.05 cm);
\fill[color4] (-3, 0) circle(0.05 cm);
\fill[color4] (-3, 1) circle(0.05 cm);
\fill[color4] (-3, 2) circle(0.05 cm);
\fill[color4] (-2, -2) circle(0.05 cm);
\fill[color4] (-2, -1) circle(0.05 cm);
\fill[color4] (-2, 0) circle(0.05 cm);
\fill[color4] (-2, 1) circle(0.05 cm);
\fill[color4] (-2, 2) circle(0.05 cm);
\fill[color4] (-1, -2) circle(0.05 cm);
\fill[color4] (-1, -1) circle(0.05 cm);
\fill[color4] (-1, 0) circle(0.05 cm);
\fill[color4] (-1, 1) circle(0.05 cm);
\fill[color4] (-1, 2) circle(0.05 cm);
\fill[color4] (0, -2) circle(0.05 cm);
\fill[color4] (0, -1) circle(0.05 cm);
\fill[color4] (0, 0) circle(0.05 cm);
\fill[color4] (0, 1) circle(0.05 cm);
\fill[color4] (0, 2) circle(0.05 cm);
\fill[color4] (1, -2) circle(0.05 cm);
\fill[color4] (1, -1) circle(0.05 cm);
\fill[color4] (1, 0) circle(0.05 cm);
\fill[color4] (1, 1) circle(0.05 cm);
\fill[color4] (1, 2) circle(0.05 cm);
\fill[color4] (2, -2) circle(0.05 cm);
\fill[color4] (2, -1) circle(0.05 cm);
\fill[color4] (2, 0) circle(0.05 cm);
\fill[color4] (2, 1) circle(0.05 cm);
\fill[color4] (2, 2) circle(0.05 cm);
\fill[color4] (3, -2) circle(0.05 cm);
\fill[color4] (3, -1) circle(0.05 cm);
\fill[color4] (3, 0) circle(0.05 cm);
\fill[color4] (3, 1) circle(0.05 cm);
\fill[color4] (3, 2) circle(0.05 cm);
\fill[color4] (4, -2) circle(0.05 cm);
\fill[color4] (4, -1) circle(0.05 cm);
\fill[color4] (4, 0) circle(0.05 cm);
\fill[color4] (4, 1) circle(0.05 cm);
\fill[color4] (4, 2) circle(0.05 cm);
\fill[color4] (5, -2) circle(0.05 cm);
\fill[color4] (5, -1) circle(0.05 cm);
\fill[color4] (5, 0) circle(0.05 cm);
\fill[color4] (5, 1) circle(0.05 cm);
\fill[color4] (5, 2) circle(0.05 cm);
\path[thicker line small arrows m=1mm in color2] (-1.9973780492810063, -0.15509132642717577) -- (-0.44352454000760066, -0.4366012477256941);
\fill[color2] (-1.9973780492810063, -0.15509132642717577) circle(0.35 cm);
\fill[white] (-1.9973780492810063, -0.15509132642717577) circle(0.175 cm);
\path[thicker line small arrows m=1mm in color2] (1.9973780492810063, 0.15909132642717577) -- (0.44352454000760066, 0.4406012477256941);
\fill[color2] (1.9973780492810063, 0.15909132642717577) circle(0.35 cm);
\fill[white] (1.9973780492810063, 0.15909132642717577) circle(0.175 cm);
  \end{tikzpicture}
          &                &
  \begin{tikzpicture}[scale=0.5]
   \fill[color4] (-5, -2) circle(0.05 cm);
\fill[color4] (-5, -1) circle(0.05 cm);
\fill[color4] (-5, 0) circle(0.05 cm);
\fill[color4] (-5, 1) circle(0.05 cm);
\fill[color4] (-5, 2) circle(0.05 cm);
\fill[color4] (-4, -2) circle(0.05 cm);
\fill[color4] (-4, -1) circle(0.05 cm);
\fill[color4] (-4, 0) circle(0.05 cm);
\fill[color4] (-4, 1) circle(0.05 cm);
\fill[color4] (-4, 2) circle(0.05 cm);
\fill[color4] (-3, -2) circle(0.05 cm);
\fill[color4] (-3, -1) circle(0.05 cm);
\fill[color4] (-3, 0) circle(0.05 cm);
\fill[color4] (-3, 1) circle(0.05 cm);
\fill[color4] (-3, 2) circle(0.05 cm);
\fill[color4] (-2, -2) circle(0.05 cm);
\fill[color4] (-2, -1) circle(0.05 cm);
\fill[color4] (-2, 0) circle(0.05 cm);
\fill[color4] (-2, 1) circle(0.05 cm);
\fill[color4] (-2, 2) circle(0.05 cm);
\fill[color4] (-1, -2) circle(0.05 cm);
\fill[color4] (-1, -1) circle(0.05 cm);
\fill[color4] (-1, 0) circle(0.05 cm);
\fill[color4] (-1, 1) circle(0.05 cm);
\fill[color4] (-1, 2) circle(0.05 cm);
\fill[color4] (0, -2) circle(0.05 cm);
\fill[color4] (0, -1) circle(0.05 cm);
\fill[color4] (0, 0) circle(0.05 cm);
\fill[color4] (0, 1) circle(0.05 cm);
\fill[color4] (0, 2) circle(0.05 cm);
\fill[color4] (1, -2) circle(0.05 cm);
\fill[color4] (1, -1) circle(0.05 cm);
\fill[color4] (1, 0) circle(0.05 cm);
\fill[color4] (1, 1) circle(0.05 cm);
\fill[color4] (1, 2) circle(0.05 cm);
\fill[color4] (2, -2) circle(0.05 cm);
\fill[color4] (2, -1) circle(0.05 cm);
\fill[color4] (2, 0) circle(0.05 cm);
\fill[color4] (2, 1) circle(0.05 cm);
\fill[color4] (2, 2) circle(0.05 cm);
\fill[color4] (3, -2) circle(0.05 cm);
\fill[color4] (3, -1) circle(0.05 cm);
\fill[color4] (3, 0) circle(0.05 cm);
\fill[color4] (3, 1) circle(0.05 cm);
\fill[color4] (3, 2) circle(0.05 cm);
\fill[color4] (4, -2) circle(0.05 cm);
\fill[color4] (4, -1) circle(0.05 cm);
\fill[color4] (4, 0) circle(0.05 cm);
\fill[color4] (4, 1) circle(0.05 cm);
\fill[color4] (4, 2) circle(0.05 cm);
\fill[color4] (5, -2) circle(0.05 cm);
\fill[color4] (5, -1) circle(0.05 cm);
\fill[color4] (5, 0) circle(0.05 cm);
\fill[color4] (5, 1) circle(0.05 cm);
\fill[color4] (5, 2) circle(0.05 cm);
\path[thicker line small arrows m=1mm in color2] (0.711138454705122, -0.5628663747426739) -- (2.212864281522734, -0.6406557441733668);
\fill[color2] (0.711138454705122, -0.5628663747426739) circle(0.35 cm);
\fill[white] (0.711138454705122, -0.5628663747426739) circle(0.175 cm);
\path[thicker line small arrows m=1mm in color2] (-0.711138454705122, 0.5668663747426739) -- (-2.212864281522734, 0.6446557441733668);
\fill[color2] (-0.711138454705122, 0.5668663747426739) circle(0.35 cm);
\fill[white] (-0.711138454705122, 0.5668663747426739) circle(0.175 cm);
  \end{tikzpicture}
  \\  \vspace{0.5cm} &&  \\
  $t=3.1$ &                & $t=7.0$
  \\ &&  \\
  \begin{tikzpicture}[scale=0.5]
   \fill[color4] (-5, -2) circle(0.05 cm);
\fill[color4] (-5, -1) circle(0.05 cm);
\fill[color4] (-5, 0) circle(0.05 cm);
\fill[color4] (-5, 1) circle(0.05 cm);
\fill[color4] (-5, 2) circle(0.05 cm);
\fill[color4] (-4, -2) circle(0.05 cm);
\fill[color4] (-4, -1) circle(0.05 cm);
\fill[color4] (-4, 0) circle(0.05 cm);
\fill[color4] (-4, 1) circle(0.05 cm);
\fill[color4] (-4, 2) circle(0.05 cm);
\fill[color4] (-3, -2) circle(0.05 cm);
\fill[color4] (-3, -1) circle(0.05 cm);
\fill[color4] (-3, 0) circle(0.05 cm);
\fill[color4] (-3, 1) circle(0.05 cm);
\fill[color4] (-3, 2) circle(0.05 cm);
\fill[color4] (-2, -2) circle(0.05 cm);
\fill[color4] (-2, -1) circle(0.05 cm);
\fill[color4] (-2, 0) circle(0.05 cm);
\fill[color4] (-2, 1) circle(0.05 cm);
\fill[color4] (-2, 2) circle(0.05 cm);
\fill[color4] (-1, -2) circle(0.05 cm);
\fill[color4] (-1, -1) circle(0.05 cm);
\fill[color4] (-1, 0) circle(0.05 cm);
\fill[color4] (-1, 1) circle(0.05 cm);
\fill[color4] (-1, 2) circle(0.05 cm);
\fill[color4] (0, -2) circle(0.05 cm);
\fill[color4] (0, -1) circle(0.05 cm);
\fill[color4] (0, 0) circle(0.05 cm);
\fill[color4] (0, 1) circle(0.05 cm);
\fill[color4] (0, 2) circle(0.05 cm);
\fill[color4] (1, -2) circle(0.05 cm);
\fill[color4] (1, -1) circle(0.05 cm);
\fill[color4] (1, 0) circle(0.05 cm);
\fill[color4] (1, 1) circle(0.05 cm);
\fill[color4] (1, 2) circle(0.05 cm);
\fill[color4] (2, -2) circle(0.05 cm);
\fill[color4] (2, -1) circle(0.05 cm);
\fill[color4] (2, 0) circle(0.05 cm);
\fill[color4] (2, 1) circle(0.05 cm);
\fill[color4] (2, 2) circle(0.05 cm);
\fill[color4] (3, -2) circle(0.05 cm);
\fill[color4] (3, -1) circle(0.05 cm);
\fill[color4] (3, 0) circle(0.05 cm);
\fill[color4] (3, 1) circle(0.05 cm);
\fill[color4] (3, 2) circle(0.05 cm);
\fill[color4] (4, -2) circle(0.05 cm);
\fill[color4] (4, -1) circle(0.05 cm);
\fill[color4] (4, 0) circle(0.05 cm);
\fill[color4] (4, 1) circle(0.05 cm);
\fill[color4] (4, 2) circle(0.05 cm);
\fill[color4] (5, -2) circle(0.05 cm);
\fill[color4] (5, -1) circle(0.05 cm);
\fill[color4] (5, 0) circle(0.05 cm);
\fill[color4] (5, 1) circle(0.05 cm);
\fill[color4] (5, 2) circle(0.05 cm);
\path[thicker line small arrows m=1mm in color2] (-0.7735862786199583, -0.388449063672759) -- (0.7382583738182118, -0.6530956305180557);
\fill[color2] (-0.7735862786199583, -0.388449063672759) circle(0.35 cm);
\fill[white] (-0.7735862786199583, -0.388449063672759) circle(0.175 cm);
\path[thicker line small arrows m=1mm in color2] (0.7735862786199583, 0.392449063672759) -- (-0.7382583738182118, 0.6570956305180558);
\fill[color2] (0.7735862786199583, 0.392449063672759) circle(0.35 cm);
\fill[white] (0.7735862786199583, 0.392449063672759) circle(0.175 cm);
  \end{tikzpicture}
          &                &
  \begin{tikzpicture}[scale=0.5]
   \fill[color4] (-5, -2) circle(0.05 cm);
\fill[color4] (-5, -1) circle(0.05 cm);
\fill[color4] (-5, 0) circle(0.05 cm);
\fill[color4] (-5, 1) circle(0.05 cm);
\fill[color4] (-5, 2) circle(0.05 cm);
\fill[color4] (-4, -2) circle(0.05 cm);
\fill[color4] (-4, -1) circle(0.05 cm);
\fill[color4] (-4, 0) circle(0.05 cm);
\fill[color4] (-4, 1) circle(0.05 cm);
\fill[color4] (-4, 2) circle(0.05 cm);
\fill[color4] (-3, -2) circle(0.05 cm);
\fill[color4] (-3, -1) circle(0.05 cm);
\fill[color4] (-3, 0) circle(0.05 cm);
\fill[color4] (-3, 1) circle(0.05 cm);
\fill[color4] (-3, 2) circle(0.05 cm);
\fill[color4] (-2, -2) circle(0.05 cm);
\fill[color4] (-2, -1) circle(0.05 cm);
\fill[color4] (-2, 0) circle(0.05 cm);
\fill[color4] (-2, 1) circle(0.05 cm);
\fill[color4] (-2, 2) circle(0.05 cm);
\fill[color4] (-1, -2) circle(0.05 cm);
\fill[color4] (-1, -1) circle(0.05 cm);
\fill[color4] (-1, 0) circle(0.05 cm);
\fill[color4] (-1, 1) circle(0.05 cm);
\fill[color4] (-1, 2) circle(0.05 cm);
\fill[color4] (0, -2) circle(0.05 cm);
\fill[color4] (0, -1) circle(0.05 cm);
\fill[color4] (0, 0) circle(0.05 cm);
\fill[color4] (0, 1) circle(0.05 cm);
\fill[color4] (0, 2) circle(0.05 cm);
\fill[color4] (1, -2) circle(0.05 cm);
\fill[color4] (1, -1) circle(0.05 cm);
\fill[color4] (1, 0) circle(0.05 cm);
\fill[color4] (1, 1) circle(0.05 cm);
\fill[color4] (1, 2) circle(0.05 cm);
\fill[color4] (2, -2) circle(0.05 cm);
\fill[color4] (2, -1) circle(0.05 cm);
\fill[color4] (2, 0) circle(0.05 cm);
\fill[color4] (2, 1) circle(0.05 cm);
\fill[color4] (2, 2) circle(0.05 cm);
\fill[color4] (3, -2) circle(0.05 cm);
\fill[color4] (3, -1) circle(0.05 cm);
\fill[color4] (3, 0) circle(0.05 cm);
\fill[color4] (3, 1) circle(0.05 cm);
\fill[color4] (3, 2) circle(0.05 cm);
\fill[color4] (4, -2) circle(0.05 cm);
\fill[color4] (4, -1) circle(0.05 cm);
\fill[color4] (4, 0) circle(0.05 cm);
\fill[color4] (4, 1) circle(0.05 cm);
\fill[color4] (4, 2) circle(0.05 cm);
\fill[color4] (5, -2) circle(0.05 cm);
\fill[color4] (5, -1) circle(0.05 cm);
\fill[color4] (5, 0) circle(0.05 cm);
\fill[color4] (5, 1) circle(0.05 cm);
\fill[color4] (5, 2) circle(0.05 cm);
\path[thicker line small arrows m=1mm in color2] (3.062140748622266, -0.608043452189312) -- (4.562285968420413, -0.6145890432135841);
\fill[color2] (3.062140748622266, -0.608043452189312) circle(0.35 cm);
\fill[white] (3.062140748622266, -0.608043452189312) circle(0.175 cm);
\path[thicker line small arrows m=1mm in color2] (-3.062140748622266, 0.612043452189312) -- (-4.562285968420413, 0.6185890432135841);
\fill[color2] (-3.062140748622266, 0.612043452189312) circle(0.35 cm);
\fill[white] (-3.062140748622266, 0.612043452189312) circle(0.175 cm);
  \end{tikzpicture}
 \end{tabular}
 \caption{\textbf{Frontal collision---C decision potential}.
  Simulation of the frontal collision described in \cref{sec:frontalcollision} under the decision function $\Phi_{C}$ using the gradient formulation. The two agents can be seen reacting to each other from a distance, steering to avoid the collision before recovering their desired direction. $L=2.0, R=1.0$. Interactive simulations available online at \href{http://rafaelbailo.com/rationalbehaviour/}{rafaelbailo.com/rationalbehaviour/}.
 }
 \label{fig:Book01CFrontal}
\end{figure}
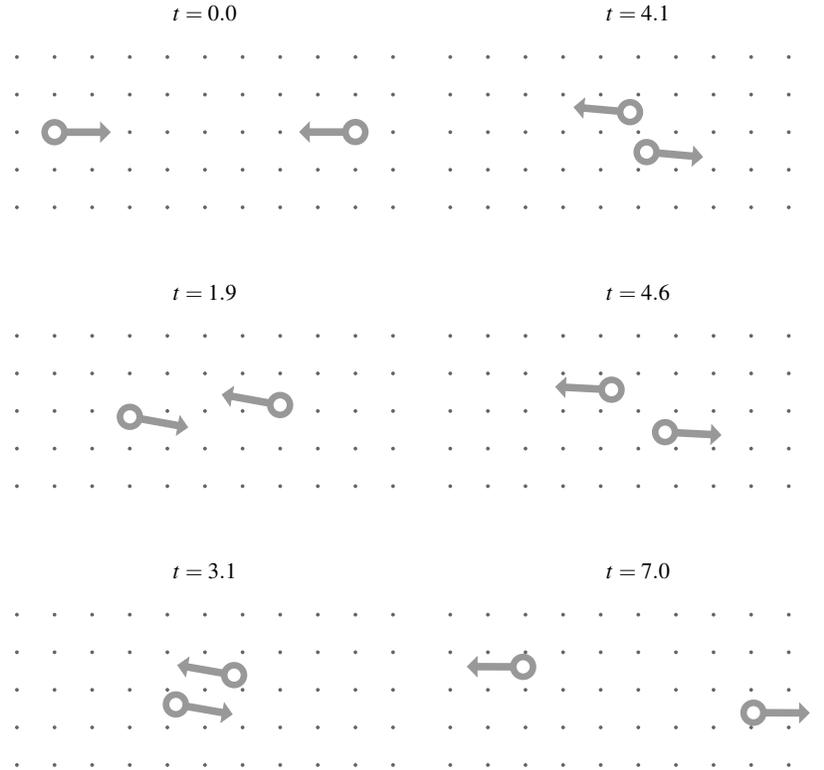

\cref{fig:comparisonphiCfrontalcollision} shows the analogous of \cref{fig:comparisonphifrontalcollision} for the new cost $\Phi_{C}$.
\cref{fig:Book01CFrontal} shows the result of a numerical simulation of the frontal collision under $\Phi_{C}$.

\subsection{Modelling Variable Speeds}

\begin{figure}
 \centering
 \sidecaption
 \begin{tabular}{ c }
  \includegraphics[width=0.70\textwidth]{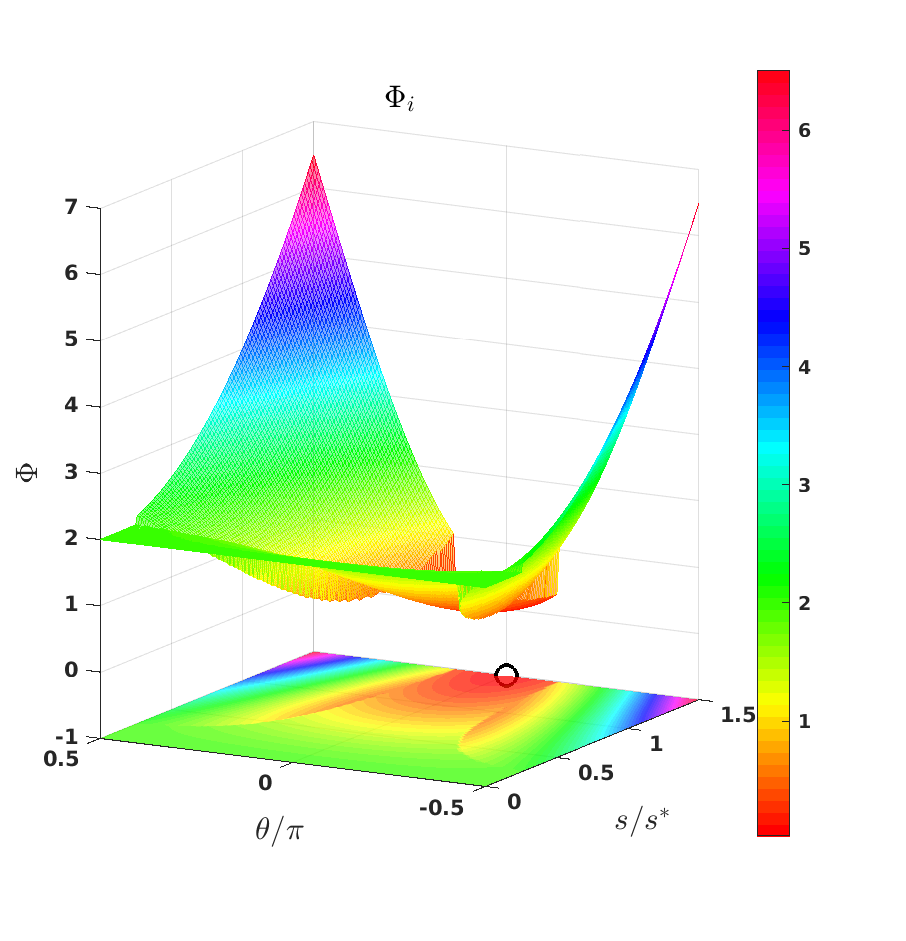}
  \\
  \includegraphics[width=0.70\textwidth]{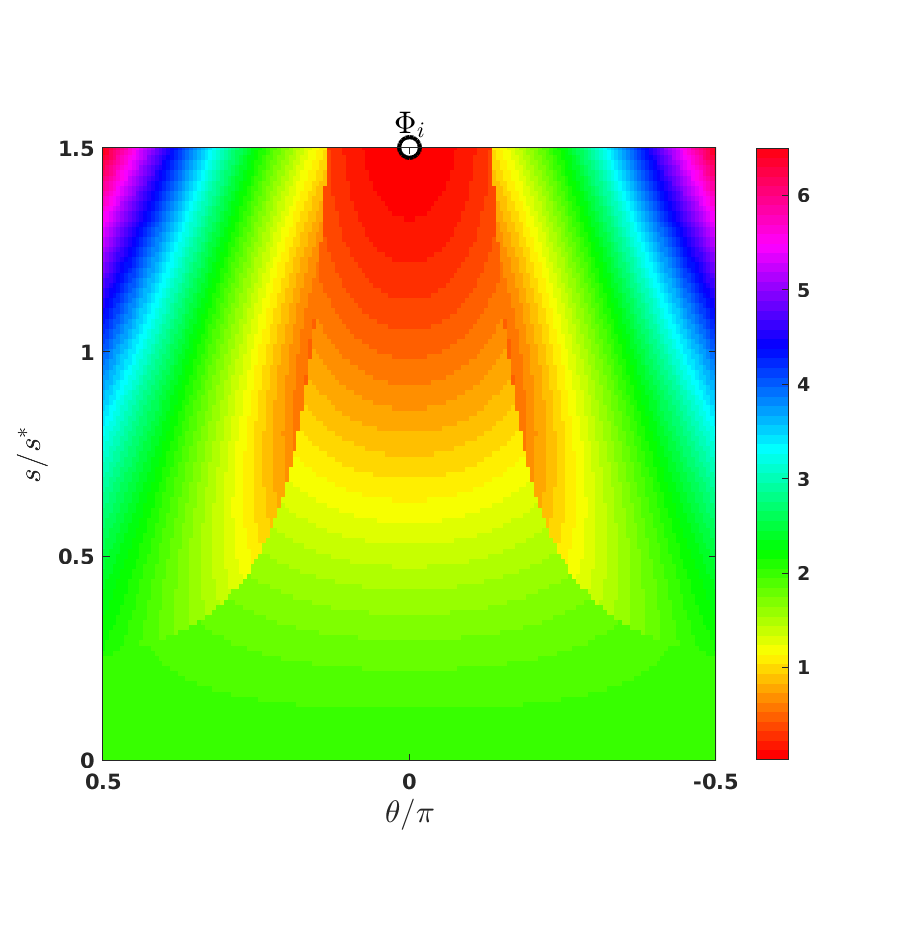}
 \end{tabular}
 \caption{
  $\Phi_i$ as a function of $s$ and $\theta$ for the frontal collision. $d=1.0, k=1.0, L=2.0, R=0.4$. Observe the marked global minimum on the boundary $s=1.5s^*$; agent $i$ will accelerate forward, in a direction close to the collision.
 }
 \label{fig:comparisonphiDspeed}
\end{figure}

\begin{figure}
 \centering
 \sidecaption
 \begin{tabular}{ c }
  \includegraphics[width=0.70\textwidth]{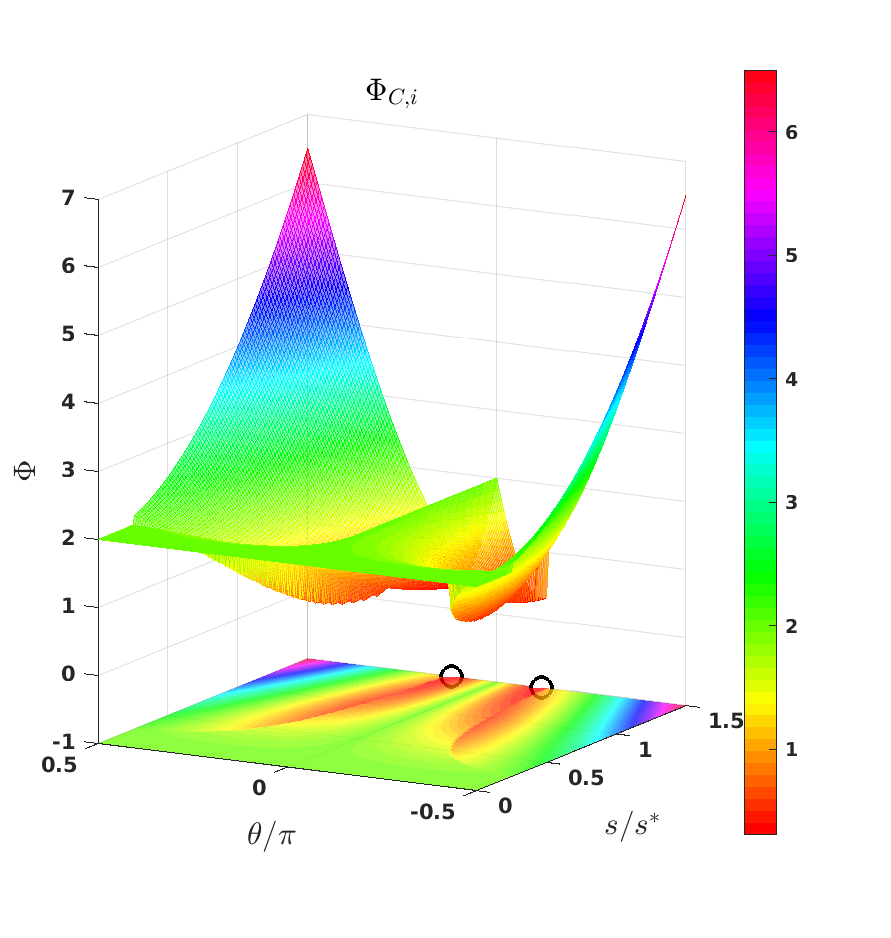}
  \\
  \includegraphics[width=0.70\textwidth]{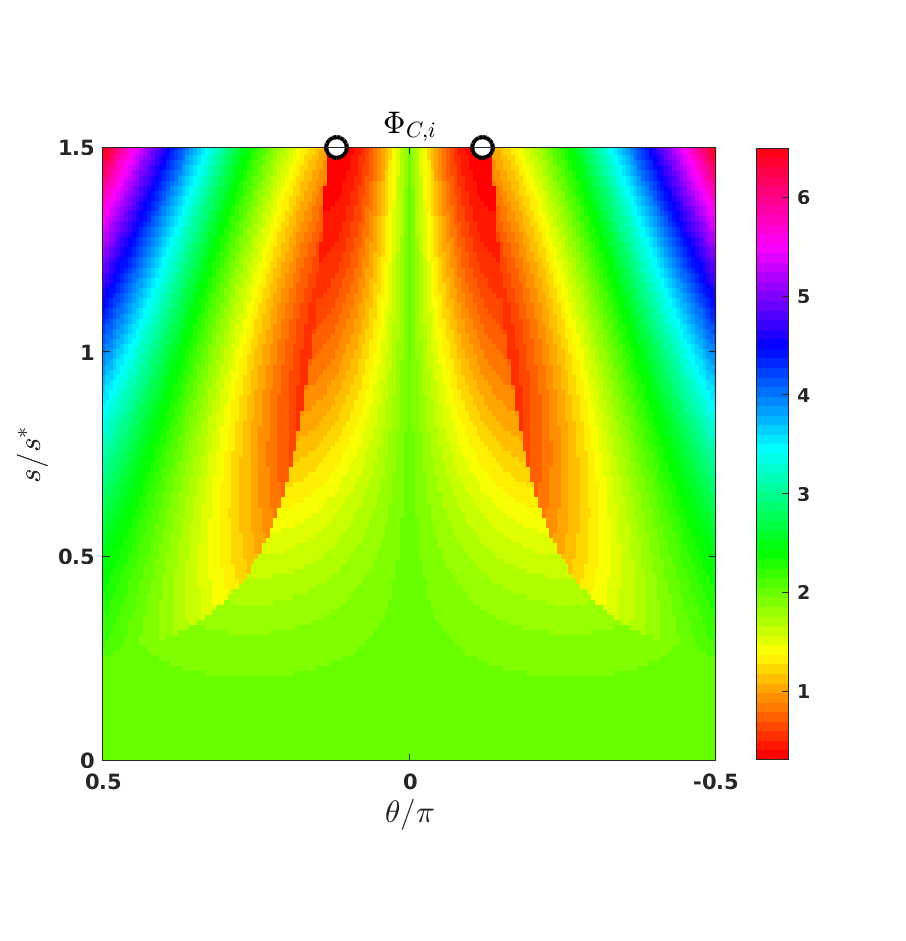}
 \end{tabular}
 \caption{
  $\Phi_{C,i}$ as a function of $s$ and $\theta$ for the frontal collision. $d=1.0, k=1.0, L=2.0, R=0.4$. Again the global minima lie on the boundary $s=1.5s^*$, leading to forward acceleration.
 }
 \label{fig:comparisonphiCspeed}
\end{figure}

To continue the generalisation of the model towards high-density regimes, we must allow for variations on the speed of pedestrians. Unfortunately the decision functions $\Phi$ and $\Phi_{C}$ have an unintended side effect on the choice of speeds.

Recall that both functions involve penalisations whenever the distance to interaction $D_i$ is less than $L$. In the case of the frontal encounter discussed above, under the gradient formulation, accelerating towards $j$ guarantees an increase of $D_i$ (and therefore a decrease of the cost), as the collision will occur closer to $j$. Hence, under the models discussed thus far, pedestrians navigating frontal collisions will accelerate towards rather than away from each other if the speed is allowed to vary. See \cref{fig:comparisonphiDspeed} and \cref{fig:comparisonphiCspeed} for a visualisation of the cost.

\begin{figure}
 \centering
 \sidecaption
 \begin{tabular}{ c }
  \includegraphics[width=0.70\textwidth]{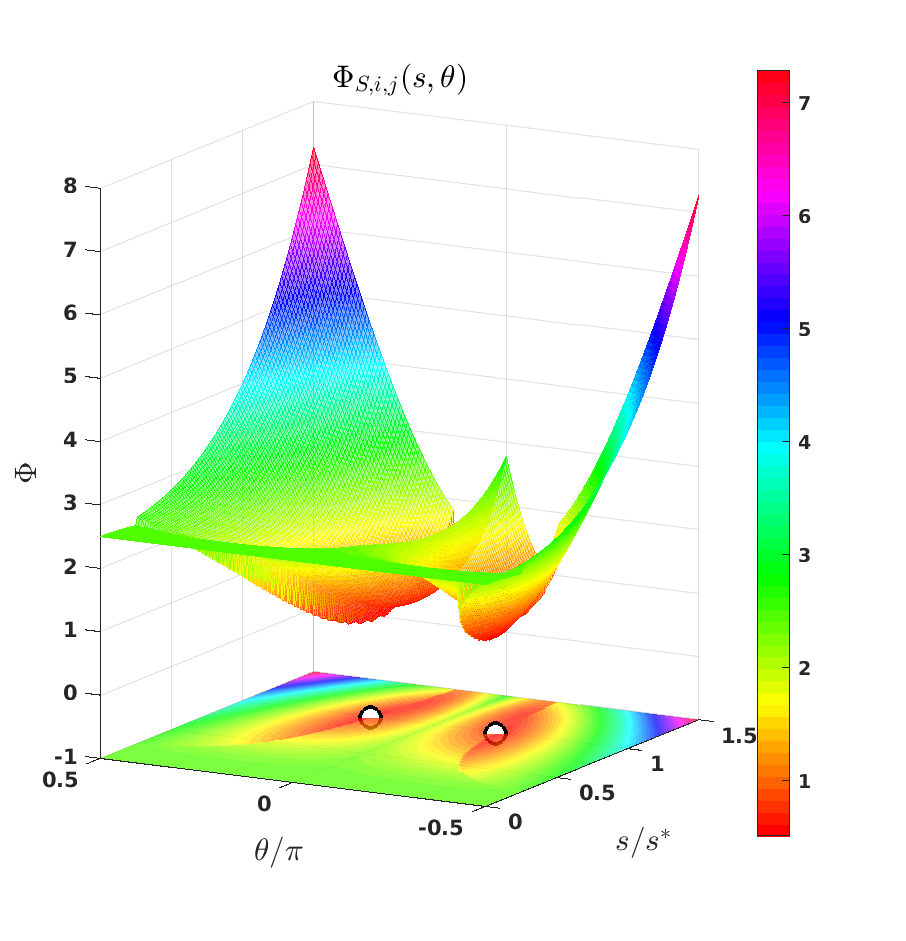}
  \\
  \includegraphics[width=0.70\textwidth]{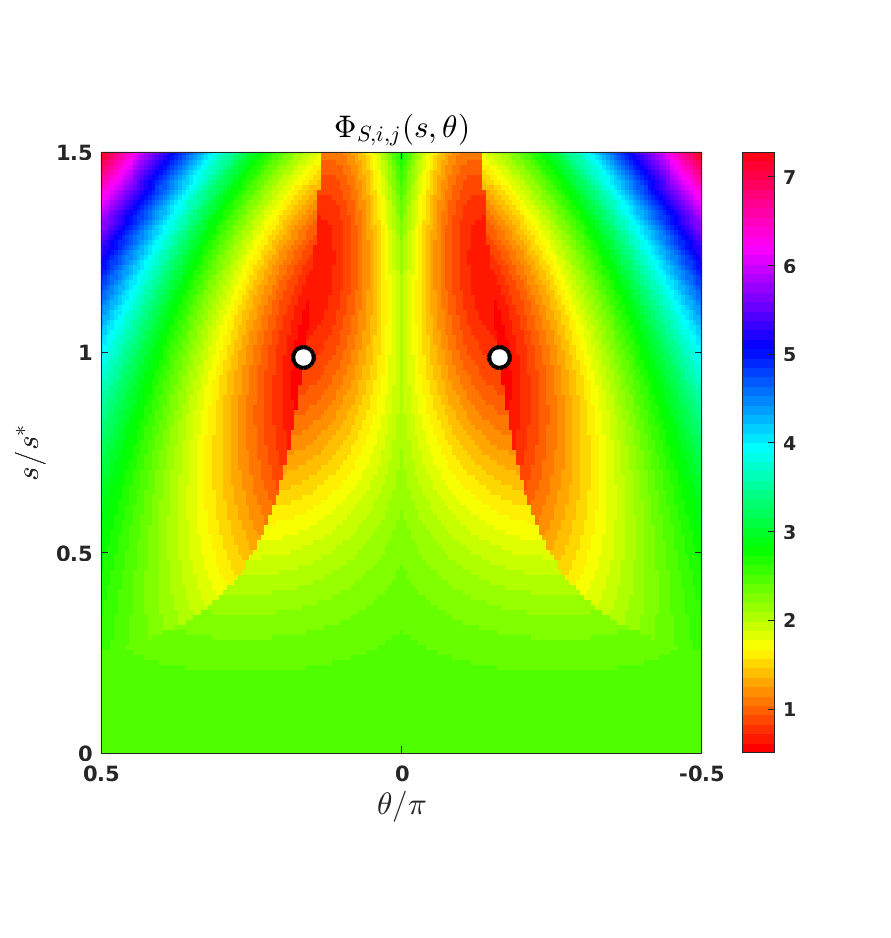}
 \end{tabular}
 \caption{
  $\Phi_{S,i}$ as a function of $s$ and $\theta$ for the frontal collision. $d=1.0, k=1.0, L=2.0, R=0.4$. Observe the marked minima have shifted away from the boundary and now sit close to $s=s^*$.
 }
 \label{fig:comparisonphiSfrontalcollision}
\end{figure}

Such peculiar acceleration suggests controlling the speed of pedestrians directly on the cost to avoid huge fluctuations away from the comfort speed $s_i^*$. We propose:
\begin{align}\label{eq:phiS}
 \Phi_{S,i}\prt{v} & =\Phi_{C,i}\prt{v}+\frac{\tilde{k}}{2}\prt{\norm{v}^2-\norm{v_i^*}^2}^2,
 \\\nonumber&=\frac{k}{2R^2}\norm{D_iC_iv-LRv_i^*}^2+\frac{\tilde{k}}{2}\prt{\norm{v}^2-\norm{v_i^*}^2}^2.
\end{align}
\cref{fig:comparisonphiSfrontalcollision} shows the modified decision potential. The minima can now be seen close to the target speed and away from the boundary.

\subsection{Environmental Coercion}\label{sec:environmental}

The modified decision potential from \eqref{eq:phiS}, $\Phi_S$, satisfactorily reflects the steering behaviour of pedestrians. As was discussed in \cref{sec:rationalbehaviour}, this decision function encodes the game-theoretical nature of human collision avoidance, where agents resolve potential collisions while attempting to remain in motion towards their target. This potential has been generalised to situations beyond the original scope (discussed in \cref{sec:frontalcollision}), allowing pedestrians to resolve collisions in high-density regimes where not every agent is able to constantly move at their desired speed.

The models as discussed thus far are only concerned with the rational avoidance of collisions by agents. However, we are yet to account for external factors that may influence the dynamics in high-density regimes. The constrains of these scenarios will be included in the gradient formulation through an additional term, the \textit{environmental coercion} $\epsilon_i$:

\begin{align}\label{eq:modelenvironmental}
 \der{x_i}{t} & =v_i,
              & \der{v_i}{t} & =-\nabla_{v}\Phi_{S,i}\prt{v_i}
 + \epsilon_i.
\end{align}

Observe the dychotomy of the \textit{decision potential} $\Phi_{S,i}$ and the \textit{environmental coercion} $\epsilon_i$. As discussed, the decision function is the principal driver of the pedestrian dynamics. The goals, strategy and overall rationality of human motion are encoded through a game of \textit{anticipation} and \textit{optimisation}. The predictive nature of the potential is made explicit through the dependence on $v$; the decision-making is always based on the predicted \textit{future} state of the agents.
Meanwhile, the environmental factor can be thought of as a higher-order correction to the model. This additional component must never dominate the dynamics and will only become significant as the pedestrian density becomes high. The term is only allowed a dependency on \textit{present} state of the agents, as it is solely a constraint due to the current density and not a rational decision process.

\subsubsection{Repulsion as Anticipation}\label{sec:repulsion}

A typical feature of high-density regimes is distance-keeping: pedestrians, particularly when in motion, maintain a safe distance away from all other agents, whether a collision is imminent or not.
This amounts to a probabilistic form of collision prevention; in avoiding close proximity, the agents are decreasing the likelihood of a collision due to a sudden change in direction by a neighbour.

\begin{figure}
 \centering
 \sidecaption
 \begin{tikzpicture}[scale=0.90]
  \begin{axis}[
    axis x line=bottom,
    axis y line=left,
    title={},
    xlabel={$r$},
    ylabel={},
    ymin=-50,
    ymax=50,
    xtick=\empty,
    ytick={0},
    yticklabels={$0$},
   ]
   \newcommand\RHOSTAR{1}
   \newcommand\repD{1}
   \newcommand\repA{1}
   \newcommand\repP{1}
   \addplot[draw opacity=0,thick,domain=0:1.1,samples=3] {10};
   \addplot[dashed,domain=0.02:1.05,samples=201] {0};
   \addplot[color1,thick,domain=0.02:1,samples=201] {\repD*exp(-\repA*x*x)*((x)^(-\repP))};
   \addplot[red,thick,domain=0.02:1,samples=201] {\repD*exp(-\repA*x*x)*((x)^(-\repP))*(-2*\repA*x-\repP*(x)^(-1))};
   \legend{,,$V\prt{r}$, $\der{V}{r}\prt{r}$}
  \end{axis}
 \end{tikzpicture}
 \caption{
  Plot of the radial potential $V$ and its derivative $f\prt{r}\defeq \der{V}{r}\prt{r}$.
 }
 \label{fig:repulsionPlot}
\end{figure}
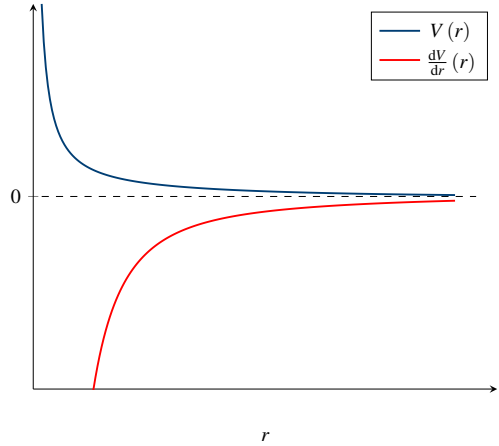

The avoidance can be modelled through the introduction of an agent-to-agent force. The environmental coercion of \eqref{eq:modelenvironmental} can be defined as a soft repulsion term:
\begin{align}\label{eq:coercionrepulsion}
 \epsilon_{f,i}\prt{x_i}\defeq &
 \sum_{j\neq i} f\prt{\norm{x_j-x_i}}\frac{{x_j-x_i}}{\norm{x_j-x_i}}.
\end{align}
An intuitive choice is to set $f$ as the derivative of a radial potential, for instance:
\begin{align}
 f\prt{r}\defeq \der{V}{r}\prt{r},\qquad V\prt{r}=D\frac{\exp\set{-ar^2}}{r^p},
\end{align}
where $a, D$ and $p$ are positive constants; see \cref{fig:repulsionPlot}.

It is important to ensure the repulsion does not dominate the dynamics, as the collision avoidance mechanism is sufficient in most cases. For instance, the frontal collision of \cref{sec:frontalcollision} can and will be resolved by steering through the choice of a suitable decision function such as \eqref{eq:phiC} or \eqref{eq:phiS}; the forces should play no role here. The function $f$ should decay rapidly to prevent middle to long distance effects, and it should be weighted by a suitably small coefficient in order to avoid sudden changes in the direction of pedestrians. Only agents that remain under close proximity during an interval of time longer than timescale of the typical collision ought to be noticeably affected by the repulsion effects.

Since the typical high-density scenario involves a large number of agents moving through a narrow geometry, it may be useful to add a similar repulsion term between each agent and the surrounding walls. Without such a term, the forces within the crowd will push agents near the boundary against the walls. This repulsion will be of a similar intensity as the agent-to-agent force, but it is imperative that it only acts on agents that approach the wall. Agents standing near a wall or moving parallel to it should experience no repulsion.

\cref{fig:Book02SCrowd} shows the result of the numerical simulation of a large crowd incorporating the repulsion effects.

\newcommand{\cropXmin}{-19}
\newcommand{\cropXmax}{19}
\newcommand{\cropYmin}{-10}
\newcommand{\cropYmax}{10}
\newcommand{\cropScale}{0.30}

\begin{figure}
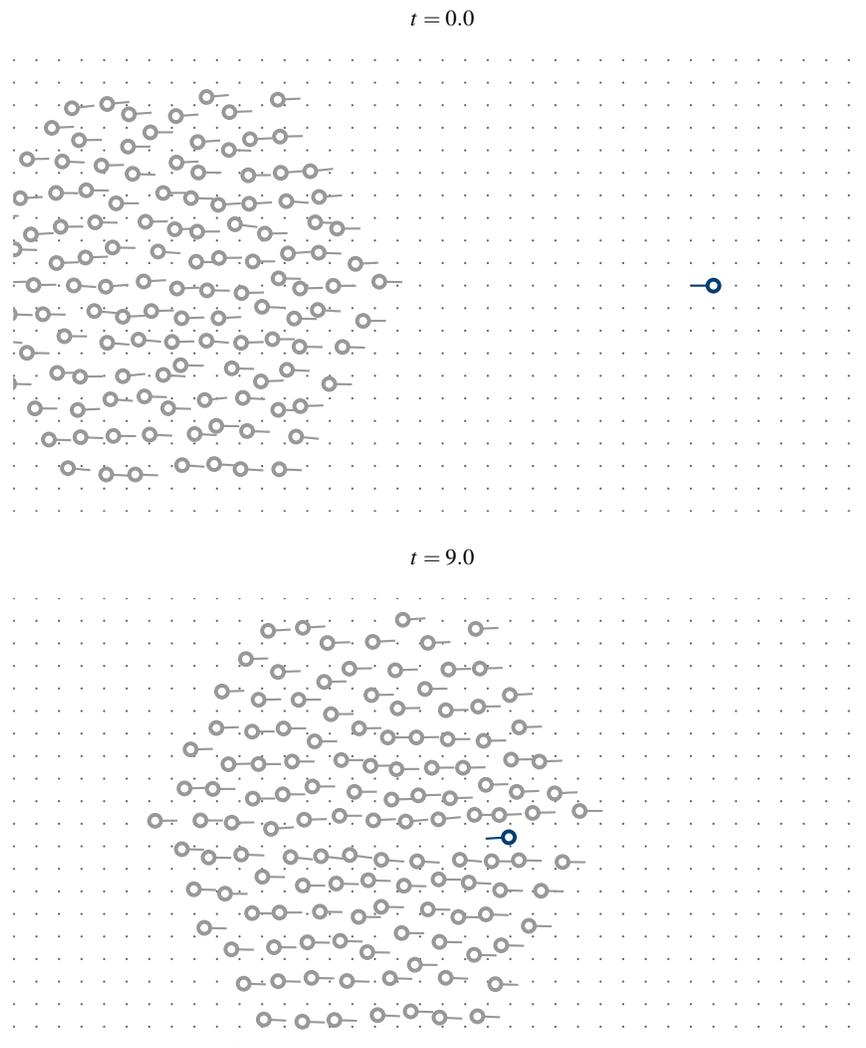

 \centering
 \sidecaption
 \begin{tabular}{ c }
  $t=0.0$
  \\  \\
  \begin{turn}{0}
   \begin{tikzpicture}[scale=\cropScale]
    \clip (\cropXmin,\cropYmin) rectangle (\cropXmax,\cropYmax);
    \input{./code/Book02SCrowd_00003000}
   \end{tikzpicture}
  \end{turn}
  \\  \\
  $t=9.0$
  \\  \\
  \begin{turn}{0}
   \begin{tikzpicture}[scale=\cropScale]
    \clip (\cropXmin,\cropYmin) rectangle (\cropXmax,\cropYmax);
    \input{./code/Book02SCrowd_00011750}
   \end{tikzpicture}
  \end{turn}
 \end{tabular}
 \caption{\textbf{Large crowd---repulsion effects}.
  Simulation of a large crowd with an incoming collision incorporating the repulsion effects of \eqref{eq:coercionrepulsion} using the gradient formulation. Agents at the front of the crowd steer to avoid the collision. After resolving the interaction, the repulsion effect causes the agents to reclaim the space that has been created on the trail of the agent, progressively returning to a homogeneous configuration. Interactive simulations available online at \href{http://rafaelbailo.com/rationalbehaviour/}{rafaelbailo.com/rationalbehaviour/}.
 }
 \label{fig:Book02SCrowd}
\end{figure}

\addtocounter{figure}{-1}
\begin{figure}
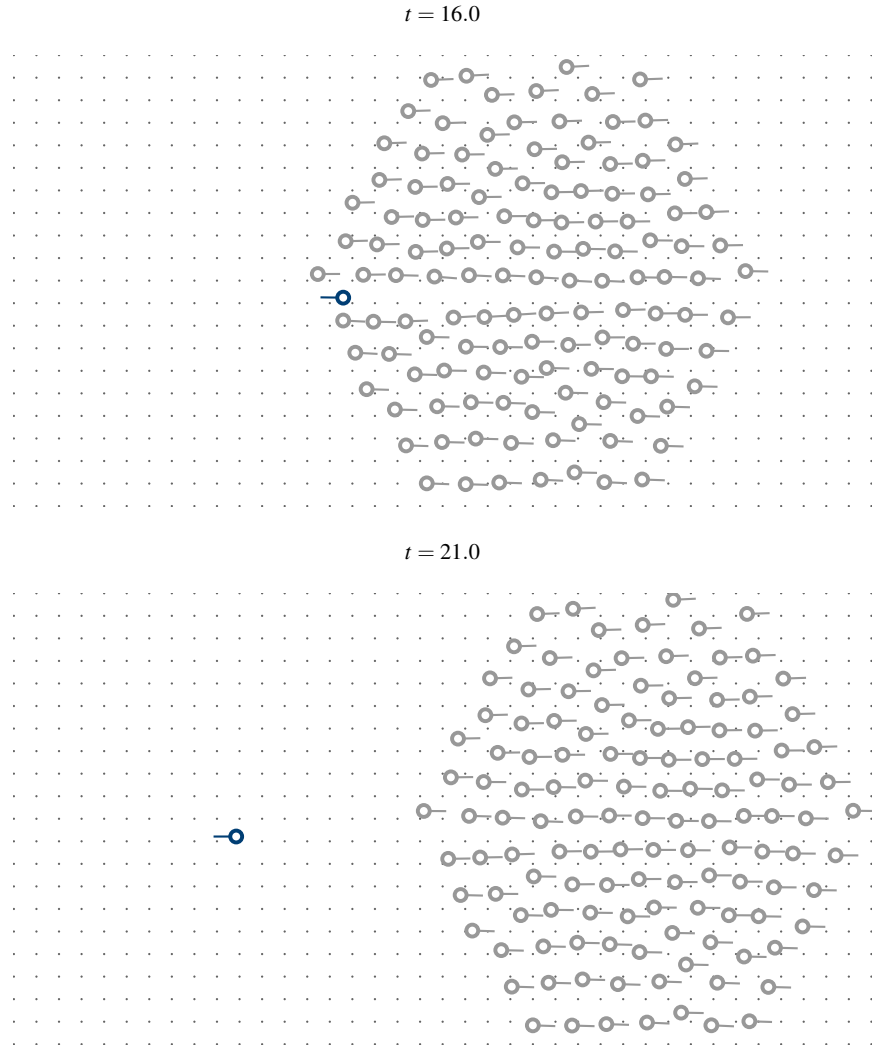

 \centering
 \sidecaption
 \begin{tabular}{ c }
  $t=16.0$
  \\  \\
  \begin{turn}{0}
   \begin{tikzpicture}[scale=\cropScale]
    \clip (\cropXmin,\cropYmin) rectangle (\cropXmax,\cropYmax);
    \input{./code/Book02SCrowd_00019020}
   \end{tikzpicture}
  \end{turn}
  \\  \\
  $t=21.0$
  \\  \\
  \begin{turn}{0}
   \begin{tikzpicture}[scale=\cropScale]
    \clip (\cropXmin,\cropYmin) rectangle (\cropXmax,\cropYmax);
    \input{./code/Book02SCrowd_00023750}
   \end{tikzpicture}
  \end{turn}
 \end{tabular}
 \caption{\textbf{Large crowd---repulsion effects (Continued)}}
\end{figure}

\subsubsection{Friction and the Fundamental Diagram}\label{sec:friction}

Another relevant behaviour in the dynamics of pedestrians is the inability to walk at full comfort speed within large crowds, even if everyone in the crowd is moving with the same velocity. The term \textit{fundamental diagram} refers to the empirical relation between crowd density and crowd speed. It has been observed that while pedestrians move at their comfort speed when moving in low densities, their movement is impaired by higher densities and their average speed is reduced; upon reaching a certain threshold the crowd is brought to a standstill.

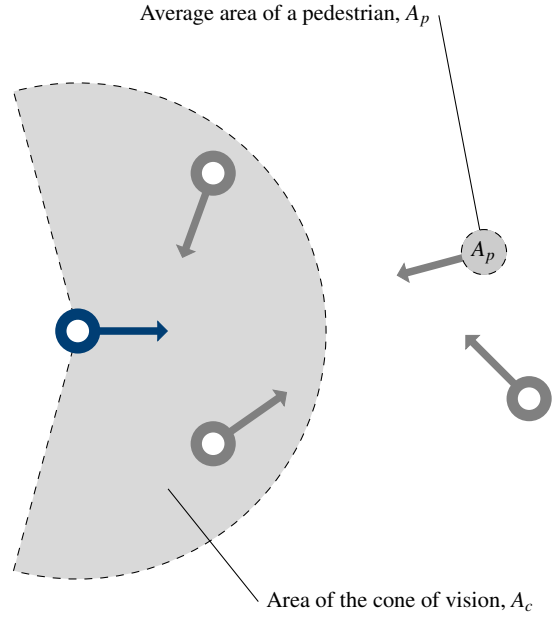
\begin{figure}
 \sidecaption
 \centering
 \begin{tikzpicture}[scale=0.6]
 \def\l{2.0}
 \def\theta{105}
 \def\r{5.5}

 \draw[dashed, fill=gray!30] (0,0) -- +({\theta}:{\r}) arc ({\theta}:{-\theta}:{\r}) -- (0,0);

 \def\xj{3}
 \def\yj{3.5}
 \def\tj{-110}

 \def\xk{3}
 \def\yk{-2.5}
 \def\tk{35}

 \def\xl{9}
 \def\yl{1.75}
 \def\tl{195}

 \def\xm{10}
 \def\ym{-1.5}
 \def\tm{135}

 \draw (2,-3.5) -- (4,-6) node[right] {Area of the cone of vision, $A_c$};
 \draw ({\xl},{\yl}) -- (8,7) node[left] {Average area of a pedestrian, $A_p$};

 \node[]() at (5.5,-8){Number of agents perceived by $i$, $N_i=2$};

 \path[thicker line small arrows m=1mm in color1] (0,0) -- (2.0,0) ;
 \fill[color1] (0,0) circle (0.50cm);
 \fill[white] (0,0) circle (0.25cm);

 \path[thicker line small arrows m=1mm in gray] (\xj,\yj) -- ({\xj+\l*cos(\tj)},{\yj+\l*sin(\tj)});
 \fill[gray] (\xj,\yj) circle (0.50cm);
 \fill[white] (\xj,\yj) circle (0.25cm);

 \path[thicker line small arrows m=1mm in gray] (\xk,\yk) -- ({\xk+\l*cos(\tk)},{\yk+\l*sin(\tk)});
 \fill[gray] (\xk,\yk) circle (0.50cm);
 \fill[white] (\xk,\yk) circle (0.25cm);

 \path[thicker line small arrows m=1mm in gray] (\xl,\yl) -- ({\xl+\l*cos(\tl)},{\yl+\l*sin(\tl)});
 \fill[gray] (\xl,\yl) circle (0.50cm);
 \fill[white] (\xl,\yl) circle (0.25cm);

 \def\shift{0}

 \draw[dashed, fill=gray!40] ({\xl+\shift},{\yl+\shift}) circle (0.50cm) node {$A_p$};

 \path[thicker line small arrows m=1mm in gray] (\xm,\ym) -- ({\xm+\l*cos(\tm)},{\ym+\l*sin(\tm)});
 \fill[gray] (\xm,\ym) circle (0.50cm);
 \fill[white] (\xm,\ym) circle (0.25cm);
\end{tikzpicture}
 \caption{
  The cone of vision. Detail of the average area of a pedestrian $A_p$ and the area of the cone of vision $A_c$. Observe that only two other agents fall within the cone of vision, $N_i=2$.
 }
 \label{fig:conevision}
\end{figure}

This speed-density coupling can be thought of a frictional force whose intensity depends on the local density of agents. Here, the \textit{local density} $\rho_i$ is the density perceived by agent $i$, which is in turn a function of the number of agents within $i$'s cone of vision. Given the \textit{number of agents perceived by $i$}, $N_i$, the \textit{average area occupied by a pedestrian} $A_p$ and the \textit{area of the cone of vision} $A_c$, the local density is simply aproximated by the ratio:
\begin{align}
 \rho_i\simeq N_i\frac{A_p}{A_c}\,,
\end{align}
see \cref{fig:conevision} for further insights.

The simplest frictional force can then be written as
\begin{align}\label{eq:coercionfriction}
 \epsilon_{\mu,i}\prt{x_i,v_i}\defeq &
 -\mu\prt{\rho_i}v_i.
\end{align}
The overall environmental coercion, a combination of the repulsion ($f$) and the friction ($\mu$) terms, becomes:
\begin{align}\label{eq:coercionoverall}
 \epsilon_i\prt{x_i,v_i}
 \defeq &
 \epsilon_{f,i}\prt{x_i}
 +
 \epsilon_{\mu,i}\prt{x_i,v_i}
 \\\nonumber= &
 \sum_{j\neq i} f\prt{\norm{x_j-x_i}}\frac{{x_j-x_i}}{\norm{x_j-x_i}}
 -\mu\prt{\rho_i}v_i.
\end{align}

\begin{figure}
 \centering
 \sidecaption
 \begin{tikzpicture}[scale=0.90]
  \begin{axis}[
    axis x line=bottom,
    axis y line=left,
    title={},
    xlabel={$\rho$},
    ylabel={$\mu\prt{\rho}$},
    ymin=0,
    ymax=50,
    xtick={1},
    xticklabels={$\rho_\textrm{max}$},
    ytick=\empty,
   ]
   \newcommand\RHOSTAR{1}
   \addplot[white,thick,domain=0:1.1,samples=3] {10};
   \addplot[color1,thick,domain=0:1,samples=201] {0.5*(\RHOSTAR) / (\RHOSTAR - x)};
   \draw[dashed] ({axis cs:1,0}|-{rel axis cs:0,0}) -- ({axis cs:1,0}|-{rel axis cs:0,1});
  \end{axis}
 \end{tikzpicture}
 \caption{
  Plot of the intensity of the frictional effect as a function of $\rho$, for a stopping density $\rho_\textrm{max}$.
  $\mu\prt{\rho}\propto\rho_\textrm{max}/\prt{\rho_\textrm{max}-\rho}$.
 }
 \label{fig:frictionPlot}
\end{figure}
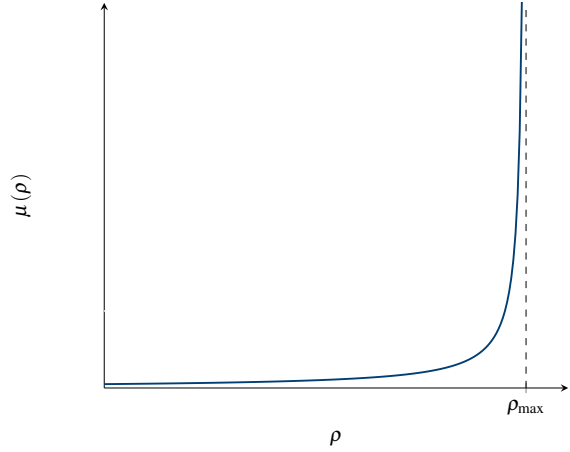

One basic possibility for the friction is $\mu\prt{\rho}\propto\rho_\textrm{max}/\prt{\rho_\textrm{max}-\rho}$, for a \textit{stopping density} $\rho_\textrm{max}$; see \cref{fig:frictionPlot}.

As in the case of the repulsion effects in \cref{sec:repulsion}, it is crucial that the friction term only dominates the dynamics in scenarios where the agent density is high. Fast, small fluctuations of the density when resolving interactions should not result in variations of the speed, as it is known that pedestrians would rather steer than deviate from their comfort speed when avoiding collisions. The effects of the new friction must only become apparent when the concentration of agents makes it impossible to avoid collisions while cruising at comfort speed.

\cref{fig:Book03FBottleneck} shows the result of the numerical simulation of a large crowd incorporating the frictional effects.

\renewcommand{\cropXmin}{-10}
\renewcommand{\cropXmax}{18}
\renewcommand{\cropYmin}{-8}
\renewcommand{\cropYmax}{8}
\renewcommand{\cropScale}{0.40}

\begin{figure}
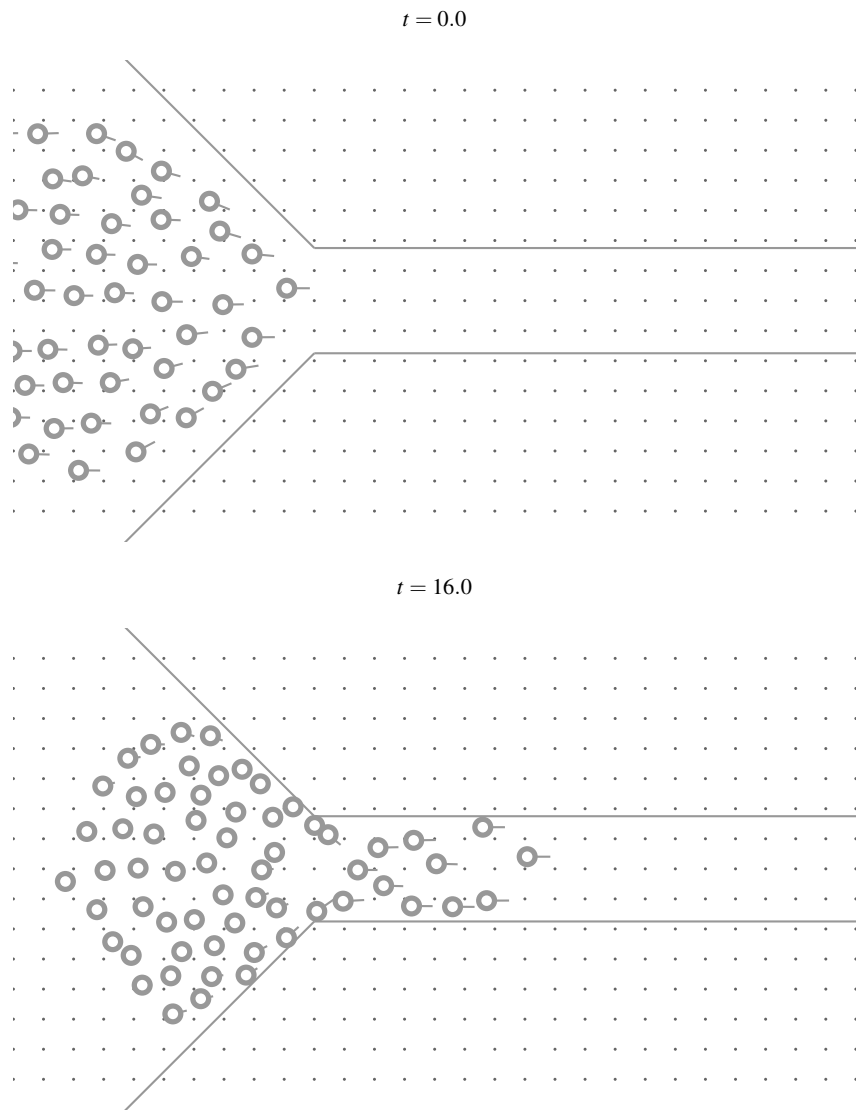

 \centering
 \sidecaption
 \begin{tabular}{ c }
  $t=0.0$
  \\  \\
  \begin{turn}{0}
   \begin{tikzpicture}[scale=\cropScale]
    \clip (\cropXmin,\cropYmin) rectangle (\cropXmax,\cropYmax);
    \input{./code/Book03FBottleneck_00010470}
   \end{tikzpicture}
  \end{turn}
  \\  \\
  $t=16.0$
  \\  \\
  \begin{turn}{0}
   \begin{tikzpicture}[scale=\cropScale]
    \clip (\cropXmin,\cropYmin) rectangle (\cropXmax,\cropYmax);
    \input{./code/Book03FBottleneck_00020860}
   \end{tikzpicture}
  \end{turn}
 \end{tabular}
 \caption{\textbf{Bottleneck---frictional effects}.
  Simulation of a large crowd navigating a bottleneck incorporating the frictional effects of \eqref{eq:coercionfriction} using the gradient formulation. Agents at the front of the crowd are able to enter the corridor unobstructed.
  As people begin to occupy the corridor the entrance quickly becomes crowded. Pedestrians waiting to enter are brought to a complete standstill until the density in front of them decreases. Interactive simulations available online at \href{http://rafaelbailo.com/rationalbehaviour/}{rafaelbailo.com/rationalbehaviour/}.
 }
 \label{fig:Book03FBottleneck}
\end{figure}

\addtocounter{figure}{-1}
\begin{figure}
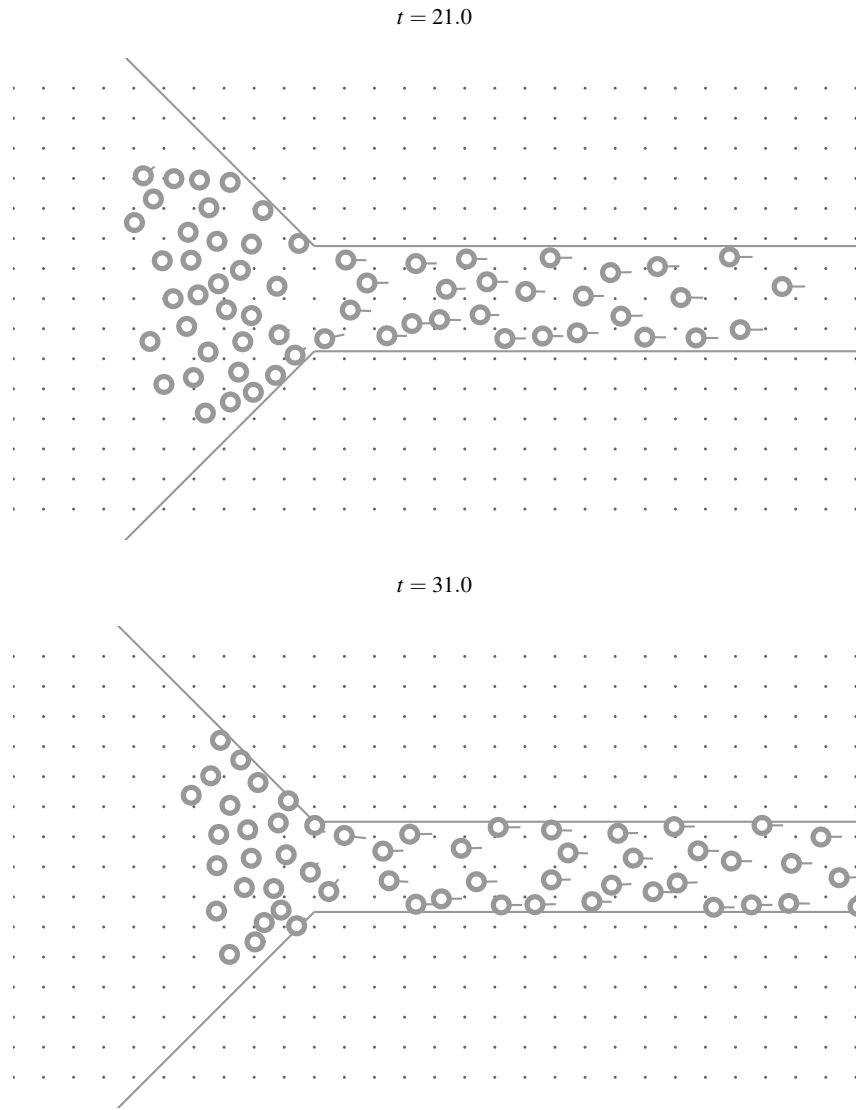

 \centering
 \sidecaption
 \begin{tabular}{ c }
  $t=21.0$
  \\  \\
  \begin{turn}{0}
   \begin{tikzpicture}[scale=\cropScale]
    \clip (\cropXmin,\cropYmin) rectangle (\cropXmax,\cropYmax);
    \input{./code/Book03FBottleneck_00031880}
   \end{tikzpicture}
  \end{turn}
  \\  \\
  $t=31.0$
  \\  \\
  \begin{turn}{0}
   \begin{tikzpicture}[scale=\cropScale]
    \clip (\cropXmin,\cropYmin) rectangle (\cropXmax,\cropYmax);
    \input{./code/Book03FBottleneck_00041880}
   \end{tikzpicture}
  \end{turn}
 \end{tabular}
 \caption{\textbf{Bottleneck---frictional effects (Continued)}}
\end{figure}

\subsection{Summary of the Modified Gradient Model}
Consider $N$ pedestrians, where agent $i$ has position $x_i$, velocity $v_i$, and target velocity $v_i^*$. The dynamics will be given by the solution to \eqref{eq:modelenvironmental}, namely:
\begin{align}
 \der{x_i}{t} & =v_i, & \der{v_i}{t}=-\nabla_{v}\Phi_{S_i}\prt{v_i} + \epsilon_i.
\end{align}
The evaluation of the \textit{decision potential} $\Phi_S$ is as follows:
\begin{enumerate}
 \item For each pair of agents $i$ and $j$, compute the heuristics $D_{i,j}$ and $C_{i,j}$ as defined in \eqref{eq:Dij} and \eqref{eq:Cij}:
       \begin{align}
        D_{i,j}= & -\frac{\prt{x_j-x_i}\cdot\prt{v_j-v_i}}{\norm{v_j-v_i}^2}\norm{v_i},
        \\C_{i,j}=&\prt{\norm{x_j-x_i}^2-\frac{\prt{\prt{x_j-x_i}\cdot\prt{v_j-v_i}}^2}{\norm{v_j-v_i}^2}}^{\frac{1}{2}}.
       \end{align}
 \item Decide whether $i$ will take $j$ into account using the conditions from \cref{sec:assumptions}:
       \begin{align}
        D_{i,j} & <L,
        \\C_{i,j}&<R,
        \\\prt{x_j-x_i}\cdot\prt{v_j-v_i}&<0,
        \\\cos(\vartheta/2)&<\frac{\prt{x_j-x_i}\cdot v_i}{\norm{x_j-x_i}\norm{v_i}}.
       \end{align}
 \item Obtain overall heuristics $D_{i}$ and $C_{i}$ as defined in \eqref{eq:Di} and \eqref{eq:Ci}:
       \begin{align}
        D_i & =D_{i,j^*},
            & C_i         & =C_{i,j^*},
            & j^*         & =\argmin_{j}\set{D_{i,j}}.
       \end{align}
 \item Use the global heuristics to construct the cost function $\Phi_S$ as defined in \eqref{eq:phiS}.
\end{enumerate}
The computation of the \textit{environmental coercion}
consists of two parts:
\begin{enumerate}
 \item The \textit{distance keeping} term, as defined in \eqref{eq:coercionrepulsion}:
       \begin{align}
        \epsilon_{f,i}\prt{x_i}\defeq &
        \sum_{j\neq i} f\prt{\norm{x_j-x_i}}\frac{{x_j-x_i}}{\norm{x_j-x_i}}.
       \end{align}
 \item The \textit{frictional} term, as given by \eqref{eq:coercionfriction}:
       \begin{align}
        \epsilon_{\mu,i}\prt{x_i,v_i}\defeq &
        -\mu\prt{\rho_i}v_i.
       \end{align}
\end{enumerate}
The \textit{overall} coercion term is the sum of the individual effects, $\epsilon_i=\epsilon_{f,i}+\epsilon_{\mu,i}$.

\

One last numerical simulation is presented in \cref{fig:CorridorHighDensity}, demonstrating the interplay between the different components of the model. Two crowds traverse a corridor in opposite directions. Initially each crowd is sparse, and agents are able to move comfortably in straight paths. As the two groups approach, interactions occur at the interface and collisions begin to be resolved. Simultaneously, as the crowds move through each other, the agent density becomes sufficiently high for the environmental constraints to manifest, leading to distance-keeping behaviour from pedestrians. Lane formation \cite{Helbing1995, Hoogendoorn2005} is observed, not as a consequence of the initial configuration of the agents but as a combined effect of the avoidance behaviours.

\renewcommand{\cropXmin}{-15}
\renewcommand{\cropXmax}{15}
\renewcommand{\cropYmin}{-8}
\renewcommand{\cropYmax}{8}
\renewcommand{\cropScale}{0.38}

\begin{figure}
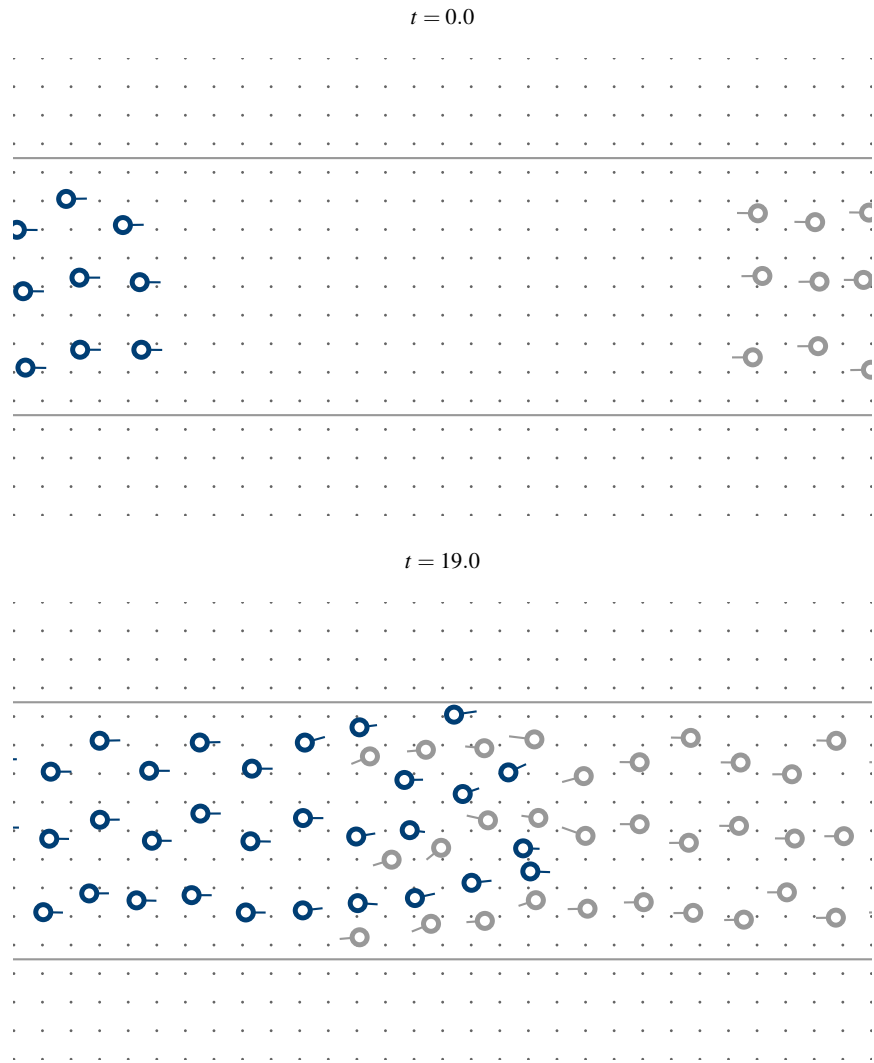

 \centering
 \sidecaption
 \begin{tabular}{ c }
  $t=0.0$
  \\  \\
  \begin{turn}{0}
   \begin{tikzpicture}[scale=\cropScale]
    \clip (\cropXmin,\cropYmin) rectangle (\cropXmax,\cropYmax);
    \input{./code/BookXXCorridor_00010000}
   \end{tikzpicture}
  \end{turn}
  \\  \\
  $t=19.0$
  \\  \\
  \begin{turn}{0}
   \begin{tikzpicture}[scale=\cropScale]
    \clip (\cropXmin,\cropYmin) rectangle (\cropXmax,\cropYmax);
    \input{./code/BookXXCorridor_00028900}
   \end{tikzpicture}
  \end{turn}
 \end{tabular}
 \caption{\textbf{Corridor---high-density setting}.
  Simulation of two dense crowds traversing a corridor in opposite directions using the gradient formulation. Agents are able to enter the corridor unobstructed at first. The initial interactions are quickly resolved through the formation of lanes, which persist in time. Interactive simulations available online at \href{http://rafaelbailo.com/rationalbehaviour/}{rafaelbailo.com/rationalbehaviour/}.
 }
 \label{fig:CorridorHighDensity}
\end{figure}

\addtocounter{figure}{-1}
\begin{figure}
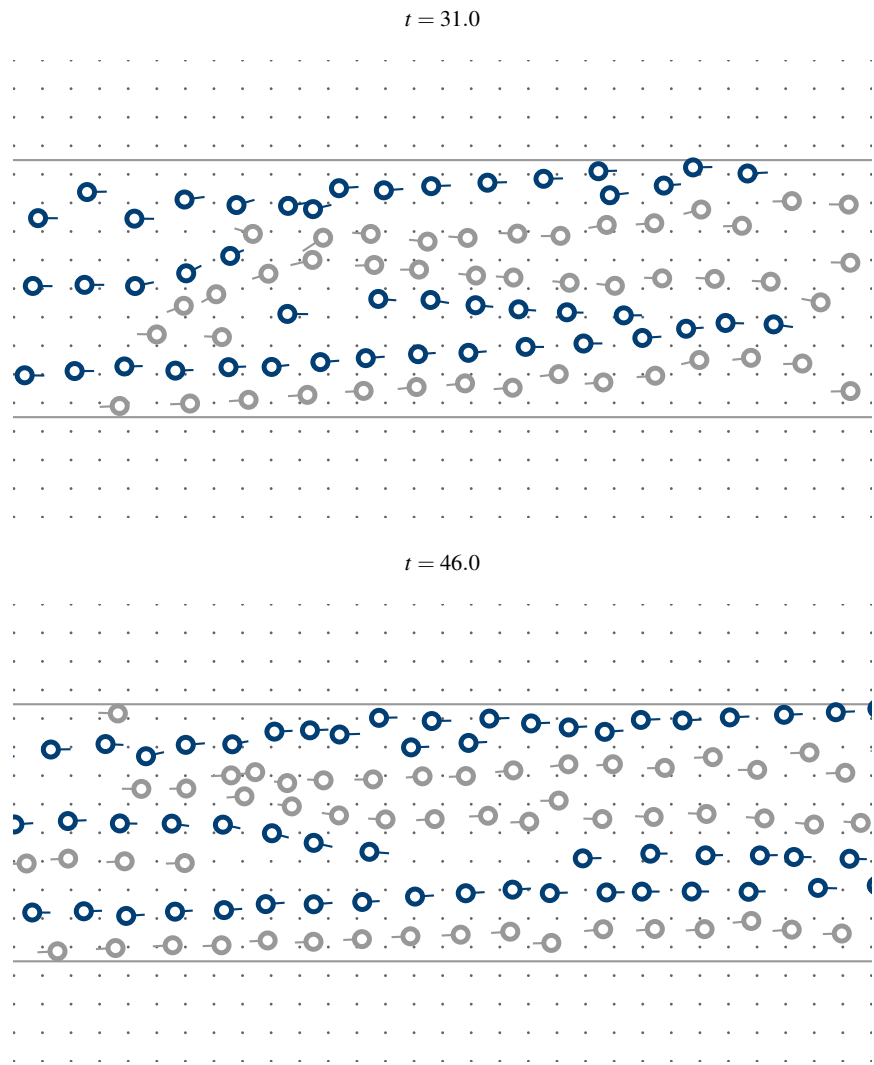

 \centering
 \sidecaption
 \begin{tabular}{ c }
  $t=31.0$
  \\  \\
  \begin{turn}{0}
   \begin{tikzpicture}[scale=\cropScale]
    \clip (\cropXmin,\cropYmin) rectangle (\cropXmax,\cropYmax);
    \input{./code/BookXXCorridor_00040900}
   \end{tikzpicture}
  \end{turn}
  \\  \\
  $t=46.0$
  \\  \\
  \begin{turn}{0}
   \begin{tikzpicture}[scale=\cropScale]
    \clip (\cropXmin,\cropYmin) rectangle (\cropXmax,\cropYmax);
    \input{./code/BookXXCorridor_00056500}
   \end{tikzpicture}
  \end{turn}
 \end{tabular}
 \caption{\textbf{Corridor---high-density setting (Continued)}}
\end{figure}

\section{Conclusion and Outlook}\label{sec:outlook}

This work has presented an individual-based model for pedestrians based on a game-theoretical principle that aims to accurately reproduce the rational behaviour of walking humans. We have explored the original formulation, which involves the use of heuristics in a decision process in order to avoid collisions. We have also explored a series of modifications to extend the validity of the model to regimes of varying characteristics.

A majority of the pedestrian models found in the literature are purely force-based. Many of these models have been applied successfully in academic and industrial settings. While, through suitable calibration, they seem to reproduce the basic principles of the dynamics and allow for the computation of crowd statistics, we feel that they do not capture the finer detail of the dynamics. As discussed previously, pedestrian motion is particularly complex, and our model seems to improve on its predecessors by faithfully replicating the rational anticipation behaviour of humans.

The immediate priority for future work will be the calibration of the parameters of the model. Each of the mechanisms described in this work involves a number of variables, many of which have a physical meaning. The currently known suitable parameters for these have been found heuristically, but a more systematic approach will be required. Different situations give rise to specific pedestrian profiles according to the context, for instance humans move differently in a train station than they do in a retail and leisure area. Such diverse dynamics, in addition to different density regimes, will require a range of calibrations of the model in order to provide adaptability. Furthermore, these calibrations should be based on real-world data to ensure fidelity.

A related line of work will involve revisiting the fundamental diagram for pedestrian dynamics in order to include it in the model in a more suitable way. The current frictional effects are too pervasive, slowing down pedestrians even when the neighbouring densities are low. Furthermore, they are prescriptive, as the form of the friction is somewhat arbitrary and should be improved. A number of machine learning techniques are now available and will be used to extract a frictional term directly from pedestrian data, rather than imposing a preconceived model.

A well-calibrated model together with an efficient implementation of the gradient formulation of \eqref{eq:modelgradient} will yield realistic live simulations. The capacity to simulate large crowds in real time will enable for the making of short-term predictions based on automated sensor data, as measurements of density and flowrates can be used to estimate an initial condition, and the model can be used to compute its evolution in time. The applications of such predictions are manyfold, allowing the anticipation and early response to undesired phenomena. Of particular interest is the optimal steering of crowds along different routes, which would be accomplished through automated signals able to adjust their information according to output from the model based on data from the crowd.

The last item of interest comprises the development of mesoscopic and macroscopic models corresponding to the dynamics of the model presented in this work. The first kinetic and hydrodynamic models derived from the original formulation appeared in \cite{Degond2013}; the comparison between these and those developed from the modified models discussed above will be relevant in understanding the properties and scales of the different components of the dynamics. Furthermore, an understanding of the correspondence between the microscopic and macroscopic scales could allow for the development of a hybrid model. This would be achieved following a \textit{level of detail} principle, where the majority of a large number of pedestrians is simulated efficiently through the macroscopic model and only the areas of particular interest are resolved at the microscopic scale.

\section*{Acknowledgements}
JAC acknowledges support by the EPSRC grant no. EP/P031587/1.
PD acknowledges support by the EPSRC grant no. EP/M006883/1, by the Royal Society and the Wolfson Foundation through a Royal Society Wolfson Research Merit Award no. WM130048. PD is on leave from CNRS, Institut de Mathématiques de Toulouse, France.
JAC and PD acknowledge support by the National Science Foundation (NSF) under Grant no. RNMS11-07444(KI-Net).

\section*{Supplementary Material}

Interactive versions of the simulations presented on Figures \ref{fig:Book01CFrontal}, \ref{fig:Book02SCrowd}, \ref{fig:Book03FBottleneck} and \ref{fig:CorridorHighDensity} are available online at 
\href{http://rafaelbailo.com/rationalbehaviour/}{rafaelbailo.com/rationalbehaviour/}
. Videos of the simulations can be found at the permanent repository 
\href{https://figshare.com/projects/Pedestrian_Models_based_on_Rational_Behaviour/38357}{figshare.com/projects/\hspace{2pt}Pedestrian\_Models\_based\_on\_Rational\_Behaviour/38357}
.

\section*{Data Statement}
No new data was generated during the course of this research.

\bibliographystyle{abbrv}
\bibliography{./bib/library}

\end{document}